\pdfoutput=1

\documentclass[iop,numberedappendix]{emulateapj}

\usepackage{natbib}
\usepackage{rotating}
\usepackage{graphicx}
\usepackage{epstopdf}
\usepackage{lineno}
\usepackage{footnote}
\usepackage[overload]{textcase}
\usepackage[colorlinks=true,linkcolor=blue,citecolor=blue,urlcolor=blue]{hyperref}

\slugcomment{To appear in the Astrophysical Journal}

\shorttitle{Variability Power Spectra of OJ\,287}
\shortauthors{Goyal et al.}

\begin{document}

\title{Stochastic modeling of multiwavelength variability of the classical BL Lac object OJ\,287 on timescales ranging from decades to hours}

\author{A.~Goyal\altaffilmark{1},
{\L}.~Stawarz\altaffilmark{1},
S.~Zola\altaffilmark{1,2},
V.~Marchenko\altaffilmark{1},
M.~Soida\altaffilmark{1},
K.~Nilsson\altaffilmark{3},
S.~Ciprini\altaffilmark{4,5},
A.~Baran\altaffilmark{2},
M.~Ostrowski\altaffilmark{1},
P.~J.~Wiita\altaffilmark{6}
% ...
}

\altaffiltext{1}{Astronomical Observatory of Jagiellonian University, ul.\ Orla 171, 30-244 Krak\'ow, Poland}
\altaffiltext{2}{Mt. Suhora Observatory, Pedagogical University, ul. \ Podchorazych 2,  Krak\'ow 30-084, Poland}
\altaffiltext{3}{Tuorla Observatory, Department of Physics and Astronomy, University of Turku, {Turku} {F-21500}, Finland}
\altaffiltext{4}{Agenzia Spaziale Italiana (ASI) Science Data Center, Roma  I-00133, Italy}
\altaffiltext{5}{Istituto Nazionale di Fisica Nucleare, Sezione di Perugia, Perugia  I-06123, Italy}
\altaffiltext{6}{Department of Physics, The College of New Jersey, 2000 Pennington Rd., Ewing, NJ 08628-0718, USA}
% ...

\email{email: {\tt arti@oa.uj.edu.pl}}

\begin{abstract}

We present the results of our power spectral density analysis for the BL Lac object OJ\,287, utilizing the {\it Fermi}-LAT survey at high-energy $\gamma$-rays,  {\it Swift}-XRT in X-rays, several ground-based telescopes and the {\it Kepler} satellite in the optical, and radio telescopes at GHz frequencies. The light curves are modeled in terms of  continuous-time auto-regressive moving average (CARMA) processes. Owing to the inclusion of the {\it Kepler} data, we were able to construct \emph{for the first time} the optical variability power spectrum of a blazar without any gaps across $\sim6$ dex in temporal frequencies. Our analysis reveals that the radio power spectra are of a colored-noise type on timescales ranging from tens of years down to months, with no evidence for breaks or other spectral features. The overall optical power spectrum is also consistent with a colored noise on the variability timescales ranging from 117 years down to hours, with no hints of any quasi-periodic oscillations. The X-ray power spectrum resembles the radio and optical power spectra on the analogous timescales ranging from tens of years down to months. Finally, the $\gamma$-ray power spectrum is noticeably different from the radio, optical, and X-ray power spectra of the source: we have detected a characteristic relaxation timescale in the {\it Fermi}-LAT data, corresponding to $\sim 150$\,days, such that on timescales longer than this, the power spectrum is consistent with uncorrelated (white) noise, while on shorter variability timescales there is correlated (colored) noise. 

\end{abstract}


\section{Introduction} 
\label{sec:intro}

Blazars are a major class of active galactic nuclei (AGN), whose total radiative energy output is dominated by the Doppler-boosted, broad-band, and nonthermal emission of relativistic jets launched by accreting supermassive black holes from the centers of massive elliptical galaxies \citep{Begelman84,DeYoung02,Meier12}. The blazar class includes BL Lacertae objects (BL Lacs) and flat-spectrum radio quasars \citep[FSRQs; see,][]{1995PASP..107..803Urry}. In the framework of the ``leptonic'' scenario for blazar emission, the radio-to-optical/X-ray segment of the continuum emission is produced by synchrotron radiation of electron-position pairs ($e^{\pm}$) accelerated up to $\sim$\,TeV energies, while the high-frequency X-ray-to-$\gamma$-ray segment is widely believed to be caused by the inverse-Comptonization of various circumnuclear photon fields (produced both internally and externally to the outflow) by the jet electrons \citep[e.g.,][]{1998MNRAS.301..451G}. Alternatively, in the ``hadronic'' scenario, the high-energy emission continuum could also be generated via protons accelerated to ultrahigh energies ($\geq$\,EeV), and producing $\gamma$-rays via either direct synchrotron emission or meson decay and synchrotron emission of secondaries from proton-photon interactions \citep[e.g.,][]{2013ApJ...768...54B}.     

Blazars display strong flux variability at all wavelengths from radio to $\gamma$-rays, on timescales ranging from decades down to hours, or even minutes. The observed flux changes are often classified broadly into three major categories, namely ``long-term variability'' (corresponding to timescales of decades to months), ``short-term variability'' (weeks to days), and intranight/day variability \citep[timescales $< 1$ day; see, e.g.,][]{1995ARA&A..33..163W,Ulrich97,2014A&ARv..22...73F}. During the last decade, special attention has been paid to catching and characterizing large-amplitude and extremely rapid (minute/hour-long) flares in the $\gamma$-ray regime, with observed intensity changes of up to even a few orders of magnitude \citep{2007ApJ...664L..71A,2011ApJ...730L...8Aleksic,Foschini11,Saito13,Rani13,2016ApJ...824L..20A}. However, these are rare, rather exceptional events while in general, the multiwavelength variability of blazar sources is of a ``colored noise'' type, meaning larger variability amplitudes on longer variability timescales with only low-level (few percent) flux changes on hourly timescales. 

More precisely, the general shapes of the power spectral densities (PSDs) of blazar light curves, which may typically be approximated to first order by a single power law $P(f)=A \, f^{-\beta}$ (where $f$ is the temporal frequency corresponding to the timescale $1/ 2 \pi f$, $A$ is the normalization constant, and $\beta>0$ is the spectral slope), indicate that the flux changes observed in given photon energy ranges are correlated over temporal frequencies \citep[][and references therein]{AG17}. So far, little or no evidence has been found for flattening of the blazar variability power spectra on the longest timescales covered by blazar monitoring programs (i.e., years and decades), even though such a flattening is expected in order to preserve the finite variance of the underlying (uncorrelated, by assumption) process triggering the variability \citep[but see in this context][]{2011A&A...531A.123K,2014ApJ...786..143S}. Breaks in the PSD slope (from $0 < \beta<2$ down to $\beta>2$ at higher temporal frequencies) reported in a few cases may, on the other hand, hint at characteristic timescales related to either a preferred location of the blazar emission zone, the relevant particle cooling timescales, or some global relaxation timescales in the systems \citep{2001ApJ...560..659K,2015ApJ...809...85F,2014ApJ...786..143S}. The detection of such break features in blazar periodograms would therefore be of primary importance for constraining the physics of blazar jets. Owing to observing constraints (e.g., weather, visibility), however, the blazar light curves from ground-based observatories are always sampled at limited temporal frequencies. This issue is particularly severe at optical and very high-energy (VHE) $\gamma$-ray energies, where generally timescales corresponding to $\sim 12$--24\,hours can hardly be probed. This difficulty has recently been surmounted in the optical range with the usage of {\it Kepler} satellite data, though only for rather limited numbers of blazars/AGN \citep{Edelson13, 2014ApJ...785...60R}.

OJ\,287 \citep[J2000.0, $\alpha {\rm=08^{h}54^{m}48\fs875}$, $\delta= +20\degr06\arcmin36\farcs64$ ; redshift $z = 0.3056$;][]{Nilsson2010}, 
is a typical example of a ``low-frequency-peaked'' BL Lac object with positive detections in the GeV and TeV photon energy ranges \citep{2010ApJ...715..429Abdo, OBrien17}. It is highly polarized in the optical band \citep[${\rm PD_{opt}} > 3\%$;][]{2011ApJS..194...19W}, and exhibits a flat-spectrum radio core with a superluminal pc-scale radio jet, both being characteristic of blazars \citep{1992ApJ...398..454Wills, 2016AJ....152...12L}. A supermassive black hole binary was claimed in the system, based on the evidence for a $\sim$\,12\,yr periodicity in its optical and radio light curves \citep{1996A&A...305L..17S, Valtonen16, Valtaoja00}; in addition, hints for a quasi-periodicity, with a characteristic timescale of $\sim 400$--800 days, have been reported for the blazar based on the decade-long optical/near-infrared and high-energy $\gamma$-ray light curves. (See, in the multiwavelength context, \citealt{2016AJ....151...54S} and \citealt{2016ApJ...832...47B}, for the most updated list of claims of quasi-periodic oscillation (QPO) detections in blazars, in general.) OJ\,287 is, in fact, one of the few blazars for which good-quality, long-duration optical monitoring data are available, dating back to circa 1896 \citep{Hudec13}. It is also one of the few blazars that have been observed by the {\it Kepler} satellite, for a continuous monitoring duration of 72 days with cadence becoming as small as 1 min. Hence, OJ\,287 is an outstanding candidate for characterizing statistical properties of optical flux changes on timescales ranging from $\sim 100$\,yr to minutes. 

\begin{table*}[ht!]
\caption{Optical observations of OJ\,287 made from ground-based observatories included in the present work}
\label{opt:LT}
\small
\begin{center}
\begin{tabular}{ccccc}\hline\hline
Data base &  Monitoring epoch (UT)     &  Filter       &   $N$      &  Reference \\     
     (1)       &   (2)                 &       (3)        &    (4)      &      (5)      \\
\hline
Harvard College Observatory$^{(a)}$ & 1900 October 4 -- 1988 November 16   & $B$    &  272    &  \citet{Hudec13}  \\
Sonneberg Observatory             & 1930 December 20 -- 1971 May 25        & $V$    & 91      &  \citet{Valtonen11} \\
Partly historical$^{(b)}$         & 1971 March 25 -- 2001 December 29      & $R$    &  3717   & \citet{Takalo94}, this work  \\
Catalina sky survey$^{(c)}$         & 2005 September 4 -- 2013 March 16      & $V$    &  606    &  this work \\
Perugia and Rome data base        & 1994 June 3 -- 2001 November 5         & $R$    &  802    &  \citet{Massaro03} \\
Shanghai Astronomical Observatory & 1995 April 19 -- 2001 December 29      & $R$    & 71      &  \citet{2003PASP..115..490Q} \\
Tuorla monitoring$^{(d)}$           & 2002 December 7 -- 2011 April 14       & $R$    &  1525   & \citet{2010MNRAS.402.2087V}, this work  \\
Krakow quasar monitoring$^{(e)}$& 2006 September 19 -- 2017 February 20  & $R$    &  1155      & \citet{2016ApJ...832...47B}, this work \\
\hline                                                                                 
\end{tabular}
\end{center}
\begin{minipage}{\textwidth}
Columns: (1) Name of the observatory/university/monitoring program; 
(2) period covered by the monitoring program (start -- end);
(3) observing filter;
(4) number of the collected data points; and
(5) references for the data (either full or partial datasets).\\ 
$^{(a)}B$-band measurements listed by \citet{Hudec13}, after applying quality cuts and removing the upper limits.\\ 
$^{(b)}$Partly displayed in Figure 3a--c of \citet{Takalo94}, converted to the $R$ band by using a constant color difference.\\ 
$^{(c)}$\texttt{http://www.lpl.arizona.edu/css/ .}\\ 
$^{(d)}$\texttt{http://users.utu.fi/kani/1m/ .}\\
$^{(e)}$\texttt{http://www.as.up.krakow.pl/sz/oj287.html .}   
\end{minipage}
\end{table*}

Here, for the first time, we present the optical PSD of OJ\,287 covering --- with no gaps --- about six decades in temporal frequency, by combining the 117\,yr-long optical light curve of the source (using archival as well as newly acquired observations with daily sampling intervals) with the {\it Kepler} satellite data. The source has also been monitored in the radio (GHz) domain with a number of single-dish telescopes, in X-rays by the space-borne {\it Swift}'s X-Ray Telescope (XRT), and in the high-energy $\gamma$-ray range with the Large Area Telescope (LAT) onboard the {\it Fermi} satellite. Here we utilize these massive datasets to derive the radio, X-ray, and $\gamma$-ray PSDs of OJ\,287, and compare them with the optical PSD.

In Section~\ref{sec:obs} we describe in more detail all the gathered data as well as the data-reduction procedures. The data analysis and the results are given in Sections~\ref{sec:carma} and \ref{sec:result}, respectively, while a discussion and our main conclusions are presented in Section~\ref{sec:conclusion}.

\section{Data acquisition and analysis: Multiwavelength light curves}
\label{sec:obs}

\subsection{High-energy $\gamma$-rays: {\it Fermi}-LAT} 
\label{sec:fermi}

We have analyzed the {\it Fermi}-LAT \citep{2009ApJ...697.1071A} data for the field containing OJ\, 287 from 2008 August 4 until 2017 February 6, and produced a source light curve between 0.1 and 300\,GeV with an integration time of 14 days. We have performed the unbinned likelihood analysis using {\it Fermi} ScienceTools {\sc 10r0p5} with {\sc p8r2\_source\_v6} source event selection and instrument response function, diffuse models {\sc gll\_iem\_v06.fits} and {\sc iso\_p8r2\_source\_v6\_v06.txt}, for the $20 ^\circ$ region centered at the blazar position, following the {\it Fermi} tutorial\footnote{\texttt{http://fermi.gsfc.nasa.gov/ssc/data/analysis/scitools/}}. The procedure starts with the selection of good data and time intervals (using the tasks {\sc ``gtselect''} and {\sc ``gtmktime''} with selection cuts {\sc evclass=128 evtype=3}), followed by the creation of an exposure map in the region of interest (ROI) with $30^\circ$ radius for each time bin (tasks {\sc ``gtltcube''}  and {\sc ``gtexpmap''}, while counting photons within zenith angle $< 90^\circ$).  

We then computed the diffuse source response (task {\sc ``gtdifrsp''}), and finally modeled the data with the maximum-likelihood method (task {\sc ``gtlike''}). In this last step, we used a model that includes OJ\,287 and 157 other point sources inside the ROI \citep[according to the third {\it Fermi}-LAT  source catalog, 3FGL;][]{Acero15}, in addition to the diffuse emission from our Galaxy ({\sc gll\_iem\_v06.fits}) and the isotropic $\gamma$-ray background ({\sc iso\_p8r2\_source\_v6\_v06.txt}) \citep{Acero16}. Furthermore, because the {\it Fermi-}LAT point-spread function has a full-wdith at half maximum of $\sim 5^{\circ}$  at energies close to $\sim$100\,MeV, variable point sources could possibly contaminate the flux of the target. We checked for variable point sources within $5^{\circ}$ of the target in the Fermi All-sky Variability Analysis (FAVA) catalog \citep{Abdollahi17}. None of the six point sources within a $5^{\circ}$ radius of the OJ\,287 position is reported to be variable in the FAVA analysis. In our modelling, we followed the standard method and fixed the photon indices and fluxes of all the point sources within the ROI other than the target at their 3FGL values. The $\gamma$-ray spectrum of OJ\,287 was modeled with a log-parabola function, according to the 3FGL shape, with the curvature (spectral parameters $a$ and $b$) and the normalization set free. We considered a successful detection to occur when the test statistic TS\,$\geq$\,10, which corresponds to a signal-to-noise ratio of $\geq 3\sigma$ \citep{2009ApJS..183...46A}. Only the data points with TS\,$\geq$\,10 have been included in the final {\it Fermi}-LAT light curve of the source, for which we have performed the timing analysis as decribed below in \S\,\ref{sec:carma}.

\subsection{X-rays: {\it Swift}-XRT} 
\label{sec:swift}

We have analyzed the archival data from the {\it Swift}-XRT \citep{2004ApJ...611.1005G}, consisting of a number of pointed observations made between 2005 May 20 and 2016 June 13. We used the latest version of calibration database ({\sc CALDB}) and version 6.19 of the {\sc heasoft} package\footnote{\texttt{http://heasarc.gsfc.nasa.gov/docs/software/lheasoft/}}. For each dataset, we used the Level 2 cleaned event files of the ``photon counting'' (PC) data-acquisition mode generated using the standard {\sc xrtpipeline} tool. 

\begin{figure*}[ht!]
\centering
\includegraphics[width=\textwidth]{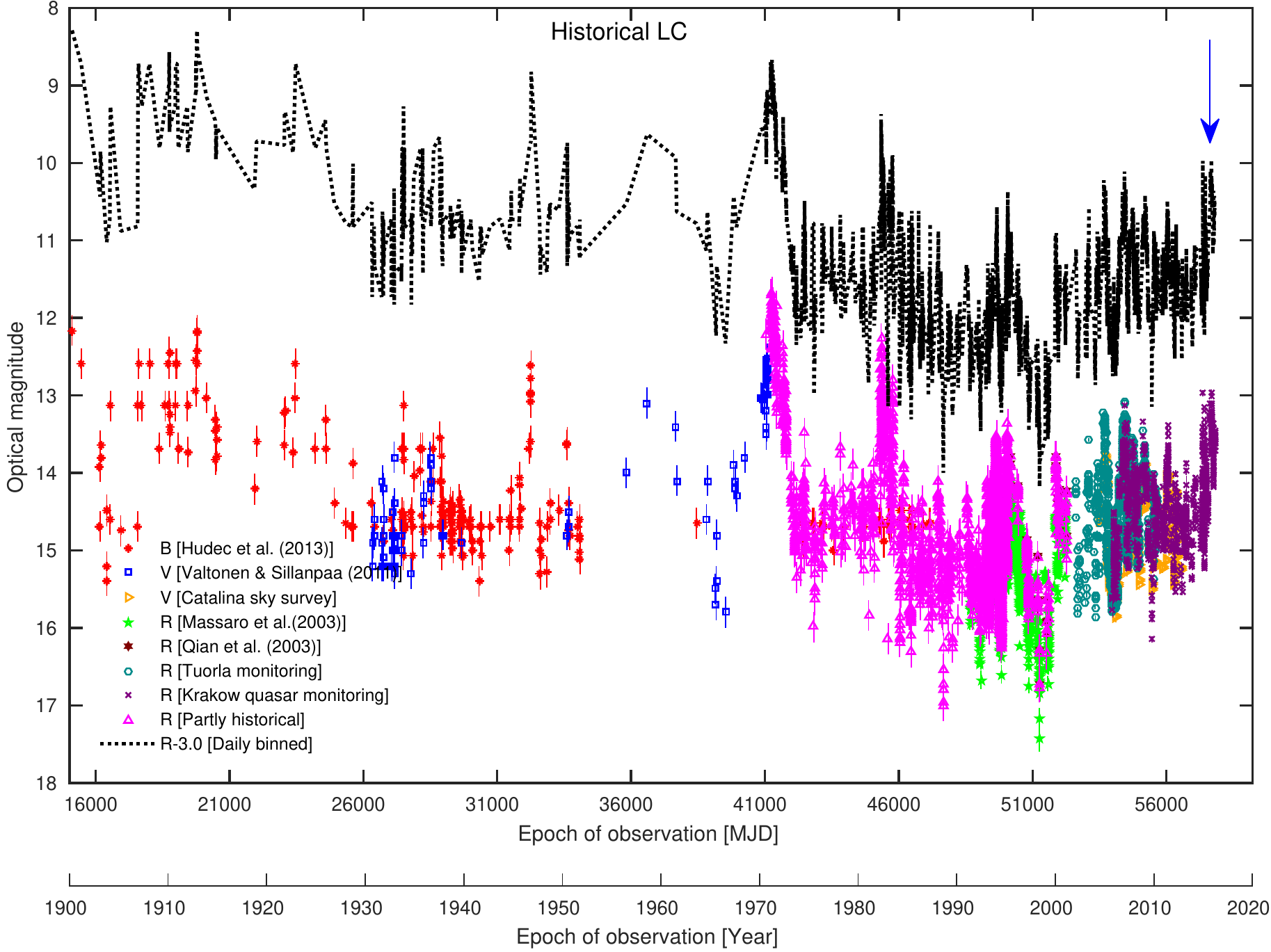}
\caption{The long-term optical light curves of OJ\,287, including the historical and also newly acquired measurements (see Section~\ref{sec:optical}). Various symbols denote the original data for the given filter used in the observations. The blue arrow indicates the maximum of the giant outburst seen in 2015 \citep[2015 December 4;][]{Valtonen16} and the black dotted line (shifted upward by 3\,mag for clarity) traces the corresponding R-band flux evolution assuming fixed color differences.}
\label{fig:1}
\end{figure*}

The source and background light curve and spectra were generated using a circular aperture with appropriate region sizes and grade filtering using the {\sc xselect} tool. The source spectra were extracted using an aperture radius of $47^{\prime\prime}$ around the source position, while a source-free region of $118^{\prime\prime}$ radius was used to estimate the background spectrum. The  ancillary response matrix was generated using the task {\sc xrtmkarf} for the exposure map generated by {\sc xrtexpomap}. All the source spectra were then binned over 20 points and corrected for the background using the task {\sc grppha}. In none of the observations did the source count-rate exceed the recommended pile-up limit for the PC mode. For each exposure, we  used routines from the X-ray data analysis software {\sc ftools} and {\sc xspec} to calculate and to subtract an X-ray background model from the data. Spectral analysis was performed between 0.3 and 10 keV by fitting a simple power law moderated by Galactic absorption with the corresponding neutral hydrogen column density fixed to $N_{\rm H, \, Gal} = 2.49 \times 10^{20}$\,cm$^{-2}$ (the task {\sc nh} in {\sc xspec}). We used the unabsorbed 0.3--10\,keV fluxes of OJ\,287 obtained in this fashion to construct the source light curve. 

\subsection{Optical: ground-based telescopes and {\it Kepler}} 
\label{sec:optical}

The long-term optical data presented in this work have been gathered from several sources and monitoring programs listed in Table~\ref{opt:LT}, including newly acquired measurements, together resulting in a {\it very long} optical light curve ranging from 1900 to 2017 February. We note that, starting from 2015 September 2, the blazar OJ\,287 has been the target of the dense multiwavelength optical monitoring campaign ``OJ287-15/16 Collaboration'' led by S.\ Zola, which was undertaken because of the predicted giant outburst in the system related to the $\sim 12$-year-long periodicity of a putative supermassive black hole binary \citep[see][and references therein]{Valtonen16, 1996A&A...305L..17S}. Details of the ``OJ287-15/16 Collaboration'' monitoring programme, including the list of observatories, are given in \citet{Valtonen16}. All the data taken from the start of 2016 through  2017 February used the skynet robotic telescope network\footnote{\texttt{http://skynet.unc.edu}} and the Mt.\ Suhora telescopes, with external observers from Greece, Ukraine, and Spain \citep[see][]{Zola16}. For these newly acquired optical data, including also the Krak\'ow quasar monitoring programme\footnote{\texttt{http://stach.oa.uj.edu.pl/kwazary/}}, data reduction was carried out using the standard procedure in the Image Reduction and Analysis Facility ({\sc iraf})\footnote{\texttt{http://iraf.noao.edu/}} software package.

The procedure starts with pre-processing of the images through bias subtraction, flat-fielding, and cosmic-ray removal. The instrumental magnitudes of OJ\,287 and the standard calibration stars listed by \citet{1996A&AS..116..403F} in the image frames were determined by aperture photometry using {\sc apphot}. This calibration was then used to transform the instrumental magnitude of OJ\,287 to a standard photometric system. Our data have quoted photometric uncertainties of $\sim 2$--5\%, arising mainly from large calibration errors in the estimated magnitudes of the stars in the field  \citep{1996A&AS..116..403F}.  A typical 0.2\,mag calibration uncertainty is assumed for $B$-band photographic magnitudes listed by \citet{Hudec13}. In the cases when during a given night the flux has been measured multiple times with different telescopes, we have averaged the measurements over the daily intervals. For the flux measurements obtained in $B$ and $V$, fixed color differences of $B-R=0.87$\,mag and $V-R=0.47$\,mag were used to convert to photometric $R$-band magnitudes in the standard Landolt photometric system \citep{1994A&AS..107..497T}. Finally, for a given $R$-band magnitude $m_R$, the $R$-band flux was derived as $3064 \times 10^{-0.4\,m_R}$\,Jy, where 3064\,Jy is the zero-point flux of the photometric system \citep{1999hia..book.....G}. The uncertainties in the $R$ fluxes were derived using standard error propagation \citep{2003drea.book.....Bevington}. The resulting long-term $R$-band historical light curve of OJ\,287 is presented in Figure~\ref{fig:1}.

OJ287 was also observed during Campaign 5 of the {\it Kepler}'s ecliptic second-life (K2) mission. The {\it Kepler} spacecraft contains a 0.95-m Schmidt telescope with a 110 square degree field of view imager having a pixel size of $4^{\prime\prime}$. It is in a heliocentric orbit currently about 0.5\,au from Earth and provides high-cadence, very-high-precision (1 part in $10^5$) photometry for rather bright stellar targets \citep{Howell14}. Campaign 5 lasted for 72 days, starting on 2015 April 27 and ending on 2015 July 10, data accumulating with both long (29.4\,min) and short  (58.85\,sec) cadences. The data are publicly available from the Barbara A. Mikulski Archive for Space Telescopes (MAST) and stored in a fits table format\footnote{\texttt{https://archive.stsci.edu/k2/}}. Timestamps are provided in Barycentric Julian Date (BJD). The long- and short-cadence data are not independent; that is why we used only the latter, which provide higher temporal resolution. In the short-cadence mode, for each timestamp a target mask is provided, while an optimal aperture is not, meaning that the light curves are not extracted. To estimate the fluxes and their uncertainties, we applied our customized scripts based on the tasks {\sc daofind} and {\sc phot} in {\sc IRAF}. We applied an aperture with a radius of 4 pixels. The background has been already subtracted during the in-house processing. The extracted light curve was subject to 4$\sigma$ clipping. The uncertainties were estimated following the recipe given in the {\sc phot} manual. 

\begin{figure}[t!]
\begin{center}
\includegraphics[width=1.1\columnwidth]{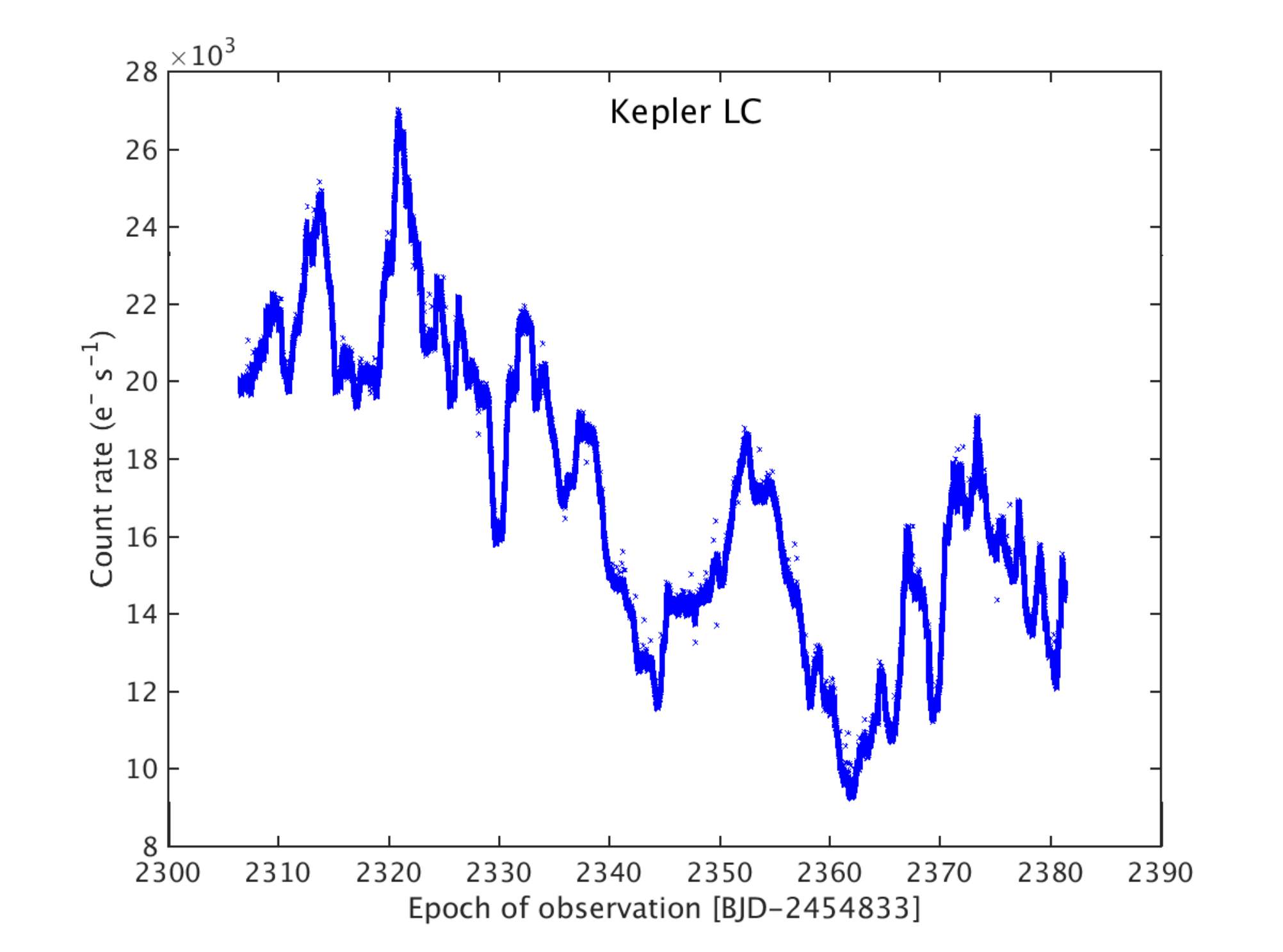}
\end{center}
\caption{The optical light curve of OJ\,287 from the {\it Kepler} satellite data analyzed in this paper (see Section~\ref{sec:optical}); the uncertainties are smaller than the point size in the figure. The thick line corresponds to the data points (overlapping in the plot), while the individual crosses (seen above and below the line) correspond to the discrepant points, mostly resulting from improper calibration of high-frequency systematics of the data.}
\label{fig:2}
\end{figure}

K2 data analysis must struggle with onboard systematics, {such as the thruster firings, reaction-wheel jitter, temperature-change-induced CCD module sensitivity, and solar-pressure-induced drift \citep{Howell14}. Most significant are the ``thruster firings,'' which cause targets to drift across the detector and usually occur at intervals of 6 hr.  The light curves are also affected at the few percent level by differential velocity aberration (DVA), which  originates  in the motion of the spacecraft in its annual orbit around the Sun. However, the DVA and thruster firings need only be of concern if they mask the intrinsic variability present in the light curve. In our case, the thruster firings add to the nominal variations at $\sim 6$ hr intervals and the DVA has relevance on multi-week timescales.  However, these rarely seem to exceed the intrinsic variations on those timescales, and we have decided not to attempt to correct for them. Unfortunately, the analysis approaches used as countermeasures for the drifts that are appropriate for detecting exoplanets  also remove real astrophysical brightness variations on the same timescales \citep[e.g.][]{2014ApJ...785...60R}. Hence, we assume that the K2 light curve found in this way is dominated by the intrinsic variability. Although some  influence from temperature changes can still be present in the light curve, this systematic could not be removed owing to the lack of calibration files provided by the archive. The K2 short-cadence light curve of OJ\,287 produced in this fashion is shown in Figure~\ref{fig:2}.

\begin{table*}[ht!]
\caption{Observational summary for the light curves of OJ\,287}
\label{tab:2}
\begin{center}
\scriptsize{
\begin{tabular}{cccccccccccccccc}\hline\hline
Data set& $N$  &    $T_{\rm obs}$    &$\Delta T_{\rm min}$ &$\Delta T_{\rm max}$   & $\Delta T_{\rm med}$   &  $\Delta T_{\rm mean}$  &  $\sigma^2_{\rm stat}$ & ${\rm log}(P_{\rm med})$ & $ {\rm log}(P_{\rm mean})$ & log(Frequency range) \\
        &      &  [yr]               & [day]             & [day]          &  [day]      & [day]   &   [rms$^2$]       & [rms$^2$ day]  &  [rms$^2$ day]    &   [day$^{-1}$]    \\
     (1)       &   (2)    &   (3)         &   (4)               &  (5)       &   (6)     &      (7)       &    (8)   & (9)    & (10)  & (11) \\        
\hline
{\it Fermi}-LAT & 199     &  8.5 &   14      &  70     & 14       &15.5       &3.0\,$10^{-1}$  &  $+$0.78   &  $+$0.81    &  $-$3.5 to $-$1.9   \\
{\it Swift}-XRT & 239     &  11  &   1       &  528    & 2.9      &18.2       &2.3\,$10^{-2}$  &  $-$0.86   &  $-$0.07    &  $-$3.6 to $-$2.1   \\
Optical (all)   & 3490    &  117 &   1       &  1728   & 1.6      &12.2       &8.9\,$10^{-2}$  &  $-$0.54   &  $+$0.33    &  $-$4.6 to $-$1.4   \\
Optical (trun.) & 3238    &  46  &   1       &  342    & 1.5      &5.2        &1.5\,$10^{-2}$  &  $-$1.37   &  $-$0.81    &  $-$4.2 to $-$0.9   \\
{\it Kepler}    & 109408  &  0.2 &  0.00068  &  0.062  & 0.00067  &0.00068    &1.2\,$10^{-5}$  &  $-$7.78   &  $-$7.78    &  $-$1.9 to $+$1.2   \\
OVRO            & 529     &  8.7 &   1       &  87     & 3.2      &6.1        &5.1\,$10^{-4}$  &  $-$2.48   &  $-$2.20    &  $-$3.5 to $-$1.5   \\
UMRAO$_{1}$     & 1364    &  33  &   2       &  468    & 6.0      &10.2       &8.8\,$10^{-4}$  &  $-$1.97   &  $-$1.74    &  $-$4.1 to $-$1.5   \\
UMRAO$_{2}$     & 1300    &  41  &   2       &  218    & 7.0      &11.6       &9.0\,$10^{-4}$  &  $-$1.89   &  $-$1.67    &  $-$4.2 to $-$1.6   \\
UMRAO$_{3}$     & 978     &  38  &   2       &  185    & 7.9      &12.4       &1.3\,$10^{-3}$  &  $-$1.69   &  $-$1.49    &  $-$4.1 to $-$1.9   \\
\hline                                                                                        
\end{tabular}
}
\end{center}
\begin{minipage}{\textwidth}
Columns: (1) Dataset; the subscripts 1, 2, and 3 (respectively) refer to the 14.5, 8.0, and 4.8\,GHz observing frequencies for the UMRAO datasets;
(2) number of data points in the observed light curve;
(3) total duration of the light curve;
(4) minimum sampling interval of the light curve;
(5) maximum  sampling interval of the light curve;
(6) median sampling interval of the light curve;
(7) mean sampling interval of the observed light curve (duration of the monitoring divided by the number of data points);
(8) mean variance of the light curve due to measurement uncertainties;
(9) median noise floor level in the PSD due to measurement uncertainty (Eq.~\ref{noise_stat});
(10) mean noise floor level in the PSD due to measurement uncertainty (Eq.~\ref{noise_stat}); and
(11) temporal frequency range covered in PSD analysis, above the median noise floor level.
\end{minipage}
\end{table*}

\subsection{Radio frequencies: UMRAO and OVRO }
\label{sec:radio}

The radio data were obtained from the University of Michigan Radio Astronomy Observatory (UMRAO) 26-m dish at 4.8, 8.0, and 14.5\,GHz, and the 40-m telescope at the Owens Valley Radio Observatory (OVRO) at 15\,GHz. The UMRAO fluxes at 4.8, 8.0, and 14.5\,GHz were typically measured weekly \citep{1985ApJ...298..296A}, from 1979 March 23 to 2012 June 15, from 1971 January 27 to 2012 May 16, and from 1974 June 20 to 2012 June 23, respectively. The OVRO light curve at 15\,GHz was sampled twice a week \citep{2011ApJS..194...29R}, during the period from 2008 January 8 to 2016 November 11.  Discussions of the corresponding observing strategies and calibration procedures can be found in \citet{1985ApJ...298..296A} for the UMRAO data, and in \citet{2011ApJS..194...29R} for the OVRO data. Figure~\ref{fig:3} (bottom panel) shows the resulting long-term radio light curves of OJ\,287, compared with the high-energy $\gamma$-ray, X-ray, and optical $R$-band light curves for the overlapping monitoring epochs. The zoomed-in version of the plot is shown in Figure~\ref{fig:4}, to highlight the {\it Swift}-XRT and {\it Fermi}-LAT data. Table~\ref{tab:2} presents a  summary of these  observational data.

\section{Power spectral density analysis: CARMA}
\label{sec:carma}

Power spectral analysis of astrophysical sources typically invokes Fourier decomposition methods, where a source light curve is represented by a sum of a set of sinusoidal signals with random phases and amplitudes, which correspond to various timescales of a source's variability in the time series \citep[e.g.,][]{TK95}. As such, the constructed PSD is a Fourier transform without the phase information. However, PSDs generated using Fourier decomposition methods can be distorted owing to {\it aliasing} and {\it red-noise leak}. Aliasing arises from the discrete sampling of a time series, while the red-noise leak appears because of the finite length of a light curve. This problem is particularly severe if a time series is not evenly sampled, as the response of a spectral window (i.e., the discrete Fourier transform of the sampling times) is in such a case unknown in the Fourier domain \citep[e.g.,][]{Deeming75}. Therefore, in order to derive reliable PSDs, an evenly sampled time series has to be obtained through a linear interpolation from an unevenly sampled data. Even though this procedure introduces interpolated data in a time series, the underlying PSD parameters can then be recovered up to a typical (mean) sampling interval of an unevenly sampled time series \citep[see the discussion in][and in particular the Appendix therein]{AG17}. 

We emphasize that the most common approaches in the literature to deal with unevenly spaced data in order to obtain PSDs using Fourier transform methods, such as the Lomb-Scargle periodogram (LSP; \citealt{Lomb76}, \citealt{Scargle82}) and the Fourier transform (FT) of the autocorrelation function \citep[ACF;][]{Edelson88}, do not reproduce correctly the slope of the underlying power spectrum. We demonstrate this in Appendix~\ref{app:A} by simulating blazar light curves using the method of \citet{Emmanoulopoulos13}. We note here that for the LSP ``least squares fitting'' method \citep[see in this context the discussion in][]{VanderPlas18}, a variability power is added at the highest frequencies --- limited by the mean sampling interval --- due to a small number of the available frequencies in the periodogram, or in other words a small number of degrees of freedom (DOF); in the periodograms derived using FT of the ACF, on the other hand, the effect of the spectral window function becomes progressively more prominent for progressively less evenly spaced data (see Appendix~\ref{app:A} for details).   

Since our aim is to characterize the variability properties of OJ\,287 over an {\it extremely broad} range of temporal frequencies ($\sim6$ decades), using the 117\,yr-long optical light curve, albeit with extremely uneven sampling, instead of standard Fourier decomposition methods here we use a certain statistical model to fit the light curve in the time domain, and thus to derive the source power spectrum.\footnote{Note that for the sparsely sampled data for OJ\,287 obtained before 1970, a linear interpolation would introduce linear trends in the timescales as long as several years, in conflict with stochastic flux changes observed on similar timescales in the much more regular monitoring conducted during the last decades.} Specifically, we use the publicly available Continuous-time Auto-Regressive Moving Average (CARMA) model\footnote{\texttt{https://github.com/brandonckelly/carma\_pack}} by \citet{Kelly14}, which is a generalized version of the first-order Continuous-time Auto-Regressive (CAR(1)) model (also known as an Ornstein-Uhlenbeck process). In the CAR(1) model, the source variability is essentially described as a damped random walk --- that is, a {\it stochastic} process with an exponential covariance function $S(\Delta t) = \sigma^2\, \exp(-|\Delta t/\tau|)$ defined by the amplitude $\sigma$ and the characteristic (relaxation) timescale $\tau$ \citep{Kelly09}. 

\begin{figure}[th!]
\begin{center}
\includegraphics[width=\columnwidth]{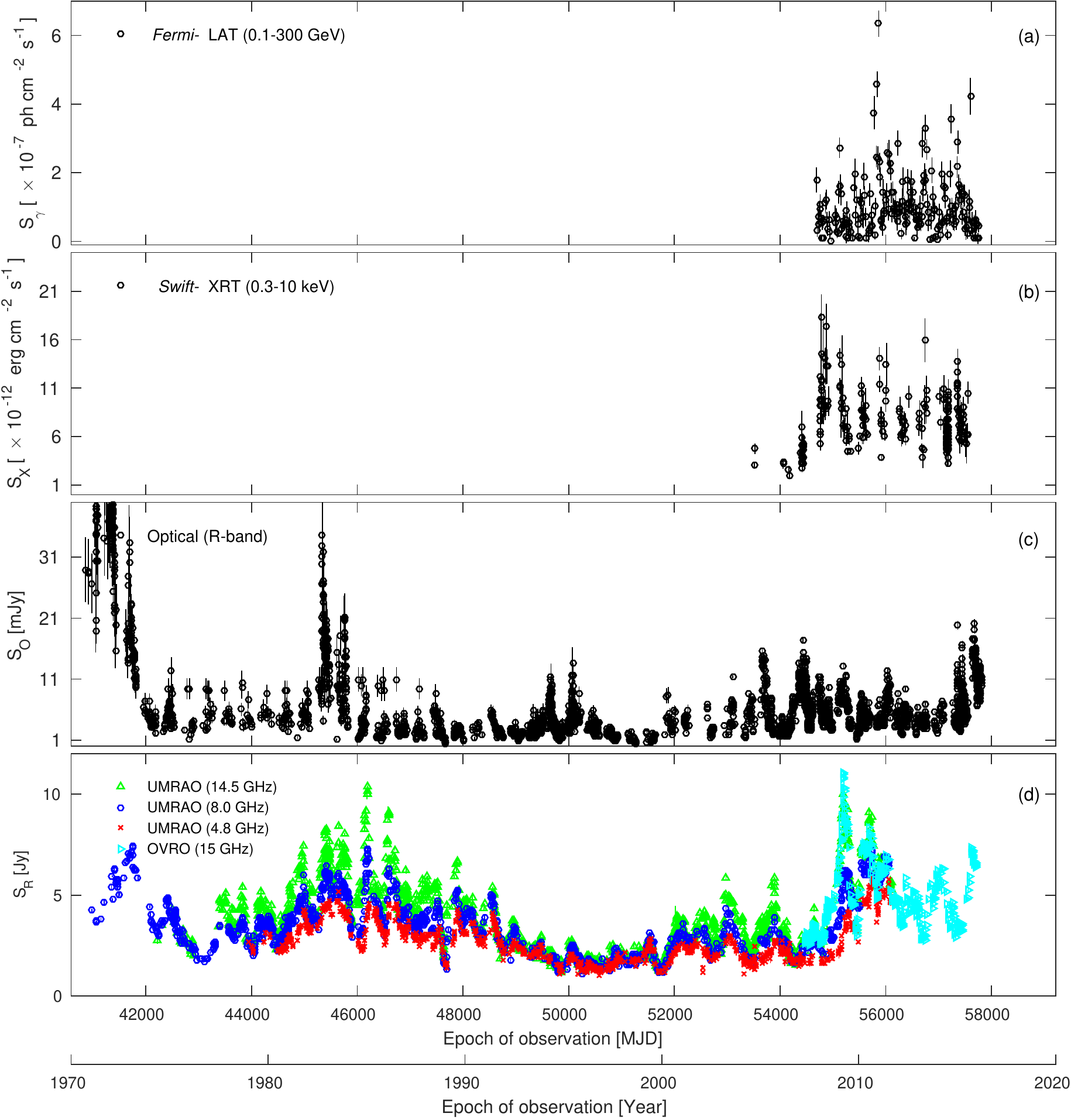}
\end{center}
\caption{The multiwavelength, long-term light curves of OJ\,287 for the monitoring epoch ranging from 1971 January until 2017 September. (a) The {\it Fermi-}LAT light curve for the energy range 0.1--300\,GeV. (b) The {\it Swift}-XRT light curve at 0.3--10\,keV. (c) The optical $R$-band light curve. (d) The 4.8--15\,GHz radio light curves, as detailed in the panel legend.}
\label{fig:3}
\end{figure}

Meanwhile, in the CARMA model, the measured time series $y(t)$ is approximated as a process defined to be the solution to the stochastic differential equation
\begin{eqnarray}
\label{eq:differential}
\frac{d^p y(t)}{dt^p} + \alpha_{p-1}\frac{d^{p-1} y(t)}{dt^{p-1}}+....+\alpha_0 y(t) \nonumber \\
= \beta_q \frac{d^{q} \epsilon(t)}{dt^{q}}+ \beta_{q-1} \frac{d^{q-1} \epsilon(t)}{dt^{q-1}}+....+\epsilon (t) \, ,
\label{eq:carma}
\end{eqnarray}
where $\epsilon(t)$ is the Gaussian (by assumption) ``input'' white noise with zero mean and variance $\sigma^2$, the parameters $\alpha_0...\alpha_{p-1}$ are the autoregressive coefficients, the parameters $\beta_1...\beta_q$ are the moving average coefficients, and finally $\alpha_p =1$ and $\beta_0 =1$ . The case with $p=1$ and $q=0$ corresponds to the CAR(1) process {(with the relaxation timescale $\tau = 2 \pi/\alpha_0$); hence, a CARMA$(p,q)$ model describes a higher-order process when compared with CAR(1).

\begin{figure}[th!]
\begin{center}
\includegraphics[width=\columnwidth]{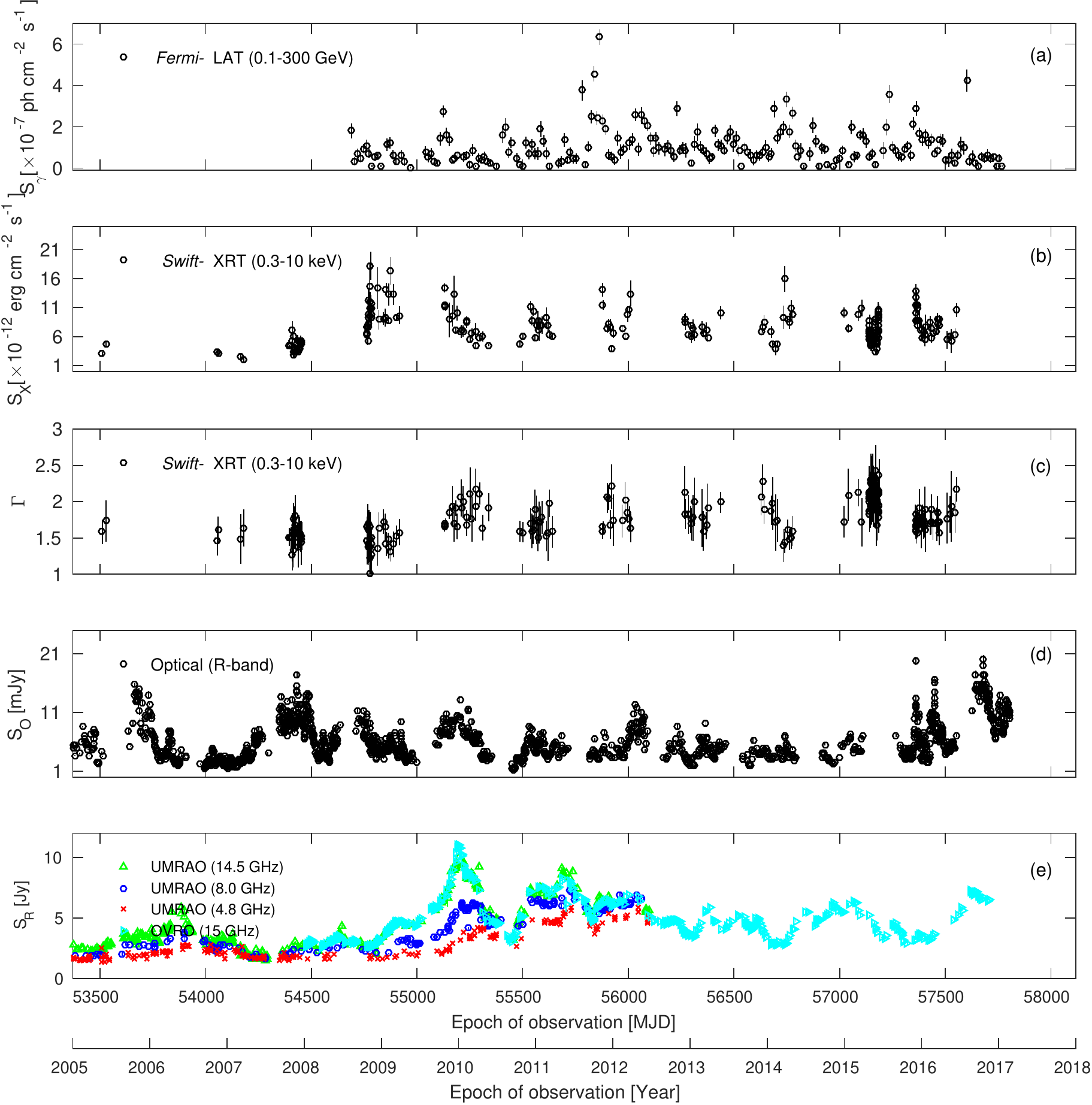}
\end{center}
\caption{A zoomed-in version of Figure~\ref{fig:3} for the last 12\,yr considered, along with a run of computed photon indices ($\Gamma$) for the {\it Swift}-XRT data (see Section~\ref{sec:swift}).}
\label{fig:4}
\end{figure}

For an in-depth discussion of the CARMA model, the reader is referred to \citet{Kelly14}. Here we only note that in this approach, for a given light curve $y(t)$, one derives the probability distribution of the (stationary) CARMA$(p,q)$ process via Bayesian inference, and in this way one calculates the corresponding power spectrum
\begin{equation}
\label{eq:psd}
P(f) = \sigma^2 \, \left|\displaystyle\sum_{j=0}^{q} \beta_j \, \left(2 \pi i f\right)^j \right|^2 \, \left|\displaystyle\sum_{k=0}^{p} \alpha_k \, \left(2 \pi i f\right)^k \right|^{-2} \, ,
\end{equation}
along with the uncertainties. \citet{Kelly14} provide the adaptive Metropolis MCMC sampler routine to obtain the maximum-likelihood estimates. The quality of the fit is assessed by standardized residuals: if the Gaussian CARMA model is correct, the residuals should form a Gaussian white-noise sequence, for which the ACF is normally distributed with mean zero and variance $1/N$, where $N$ is the number of data points in the measured time series. Importantly, since here a light curve is directly \emph{modeled} in the time domain --- unlike in a PSD analysis using the standard FT methods --- the resulting PSD shape should, in principle, be free from ``uneven sampling'' effects.

Here we employ the generalized ``corrected'' Akaike Information Criterion \citep[AICc;][]{Hurvich89}, which is based on the maximum-likelihood estimate of the model parameters, including penalizing against overfitting due to the model complexity for finite sample sizes, to chose the order of the CARMA$(p,q)$ process, although other criteria could be applied in this context as well. With such, models having AICc values $< 2$ can be considered as sufficiently close to the null hypothesis, while models having AICc values $> 10$ are not supported \citep{Burnham04}. The CARMA software package {from \citet{Kelly14} that we employed finds the maximum-likelihood estimates of the model parameters by running 100 optimizers with random initial sets of model parameters, and then selects the order $(p,q)$ that minimizes the AICc.     

Finally, we note that the noise floor level in the derived PSD (Eq. \ref{eq:psd}) resulting from statistical fluctuations caused by measurement errors, is calculated as
\begin{equation}
P_{\rm stat} = 2\, \Delta t \, \sigma_{\rm stat}^2 \, ,
\label{noise_stat}
\end{equation}
where $ \Delta t$ is the sampling interval and $\sigma_{\rm stat}^2= \sum_{j=1}^{j=N} \Delta y (t_j)^2 / N$ is the mean variance of the measurement uncertainties in the flux values $y(t_j)$ in the observed light curve at times $t_j$.

\begin{figure*}
\begin{center}
\includegraphics[width=1.0\textwidth]{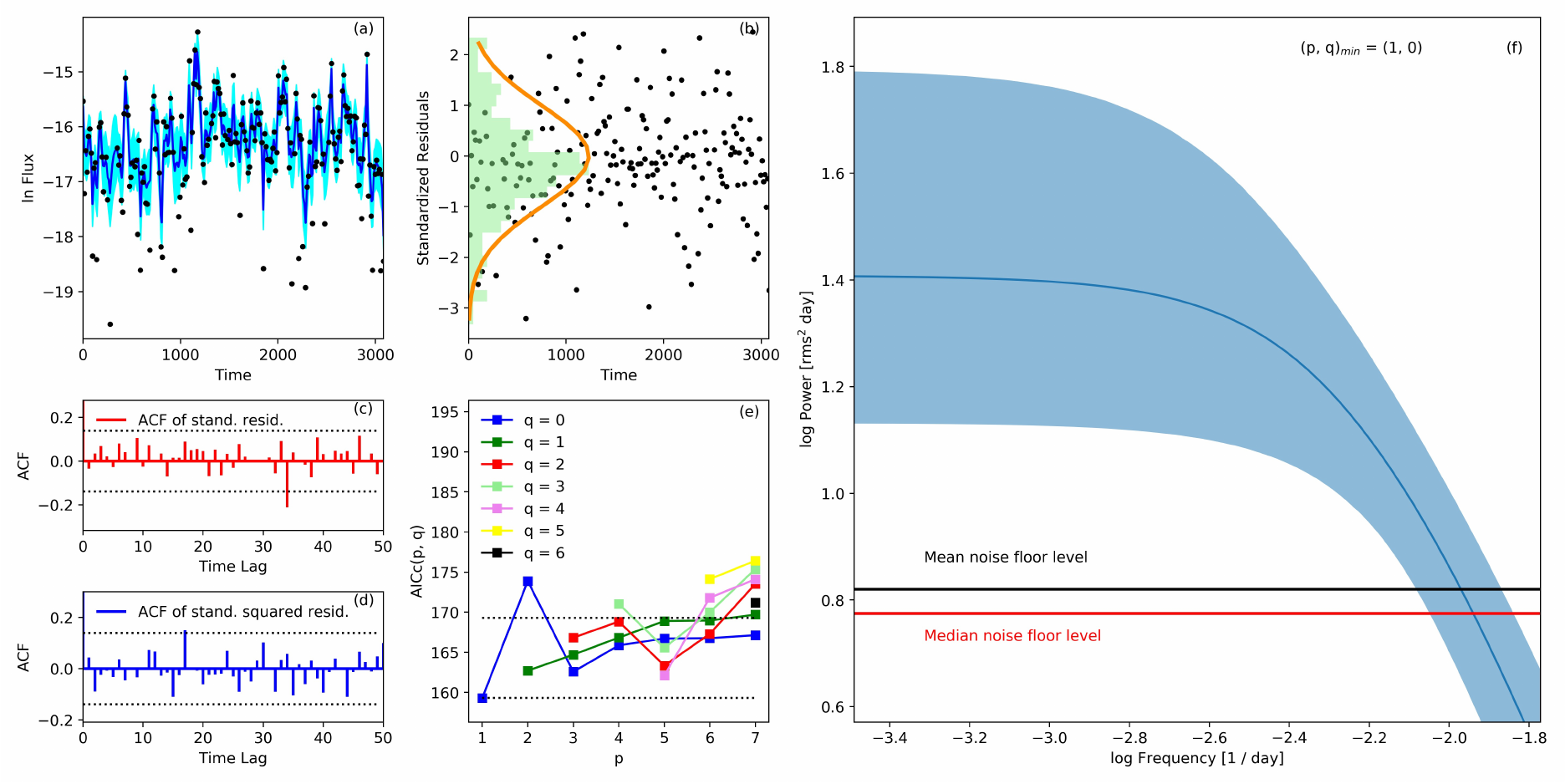}
\end{center}
\caption{(a) The {\it Fermi}-LAT light curve of OJ\,287 (black points), along with the modeled values based on the best-fitting CARMA(1,0) process selected according to the minimum AICc value (blue curve). (b) Standardized residuals (black points) and their distribution (green histogram), compared with the expected normal distribution (orange curve). (c) The corresponding ACF, compared with the 95\% confidence region assuming a white-noise process (dashed horizontal lines). (d) The corresponding squared ACF, compared with the 95\% confidence region assuming a white-noise process (dashed horizontal lines). (e) The AICc values for various CARMA$(p,q)$ models of the order $p \leq 7$ and $q< p$; the minimum AICc value is achieved for $(p=1, q=0)$, but the region between the horizontal dotted lines denotes the models which are statistically indistinguishable. (f) The CARMA(1,0) model PSD of the {\it Fermi}-LAT light curve, along with the $2\sigma$ confidence region (blue area), as well as the mean and median noise floor levels (black and red lines, respectively).}
\label{fig:5}
\end{figure*}

\begin{figure*}
\begin{center}
\includegraphics[width=1.0\textwidth]{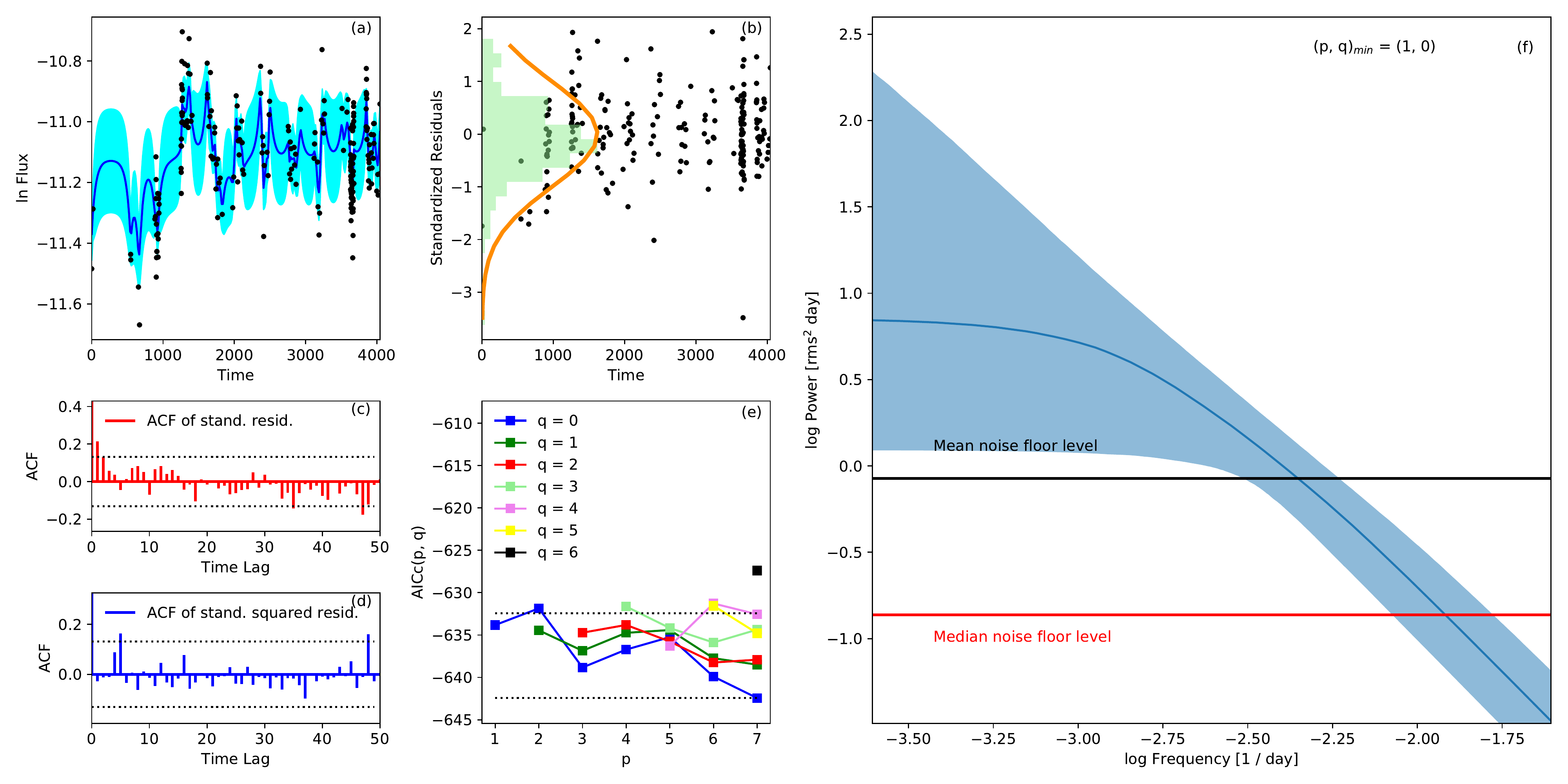}
\end{center}
\caption{As in Figure~\ref{fig:5}, but for the {\it Swift}-XRT data, very well fit by the CARMA(1,0) model, although the CARMA(3,1) and CARMA(4,0) models are nominally favored.}
\label{fig:6}
\end{figure*}

\begin{figure*}
\begin{center}
\includegraphics[width=1.0\textwidth]{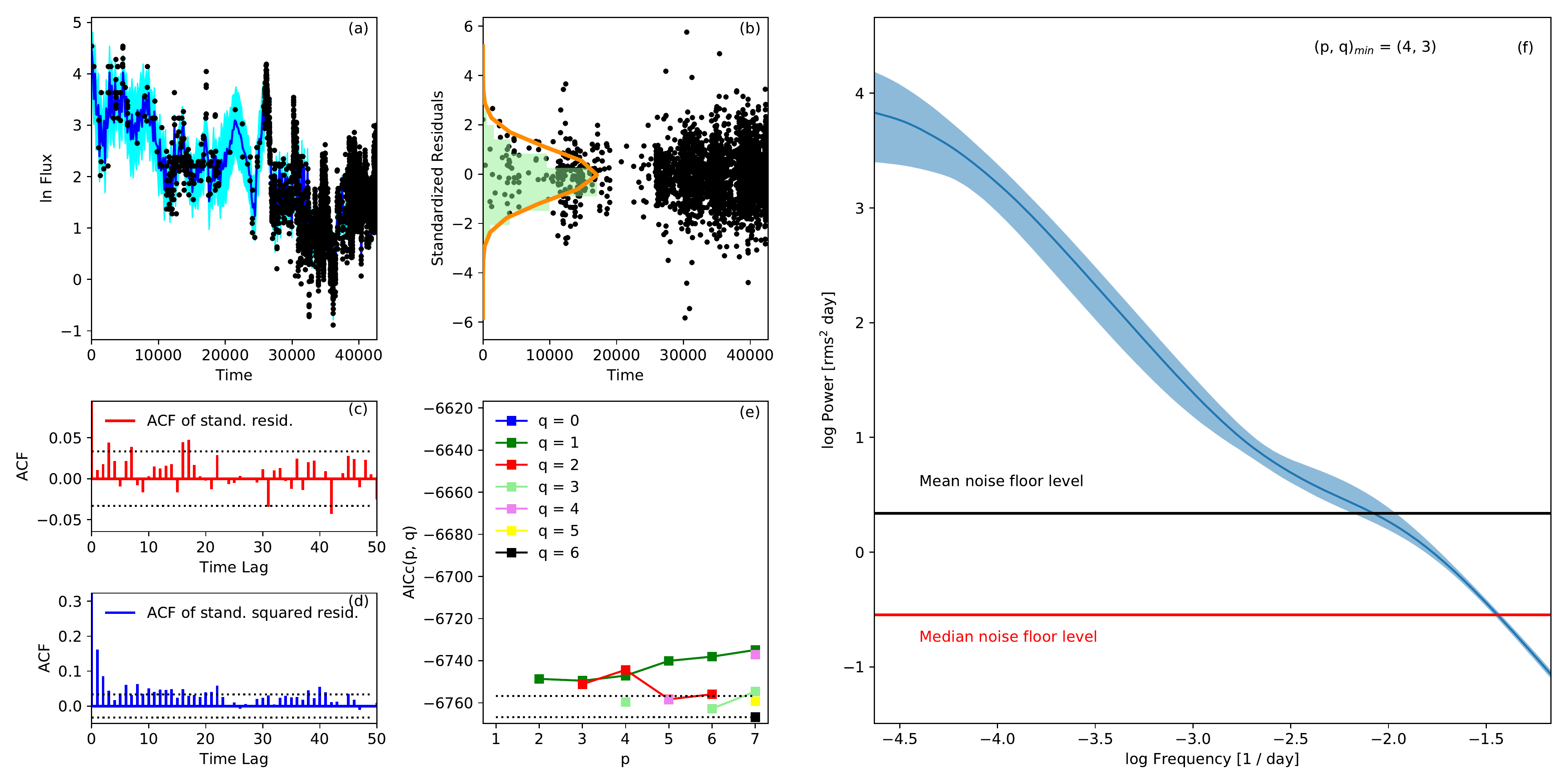}
\end{center}
\caption{As in Figure~\ref{fig:5}, but for the entire optical data (optical(all); Table~\ref{tab:2}), which is very well fit by the CARMA(4,3) model.}
\label{fig:7}
\end{figure*}

\begin{figure*}
\begin{center}
\includegraphics[width=1.0\textwidth]{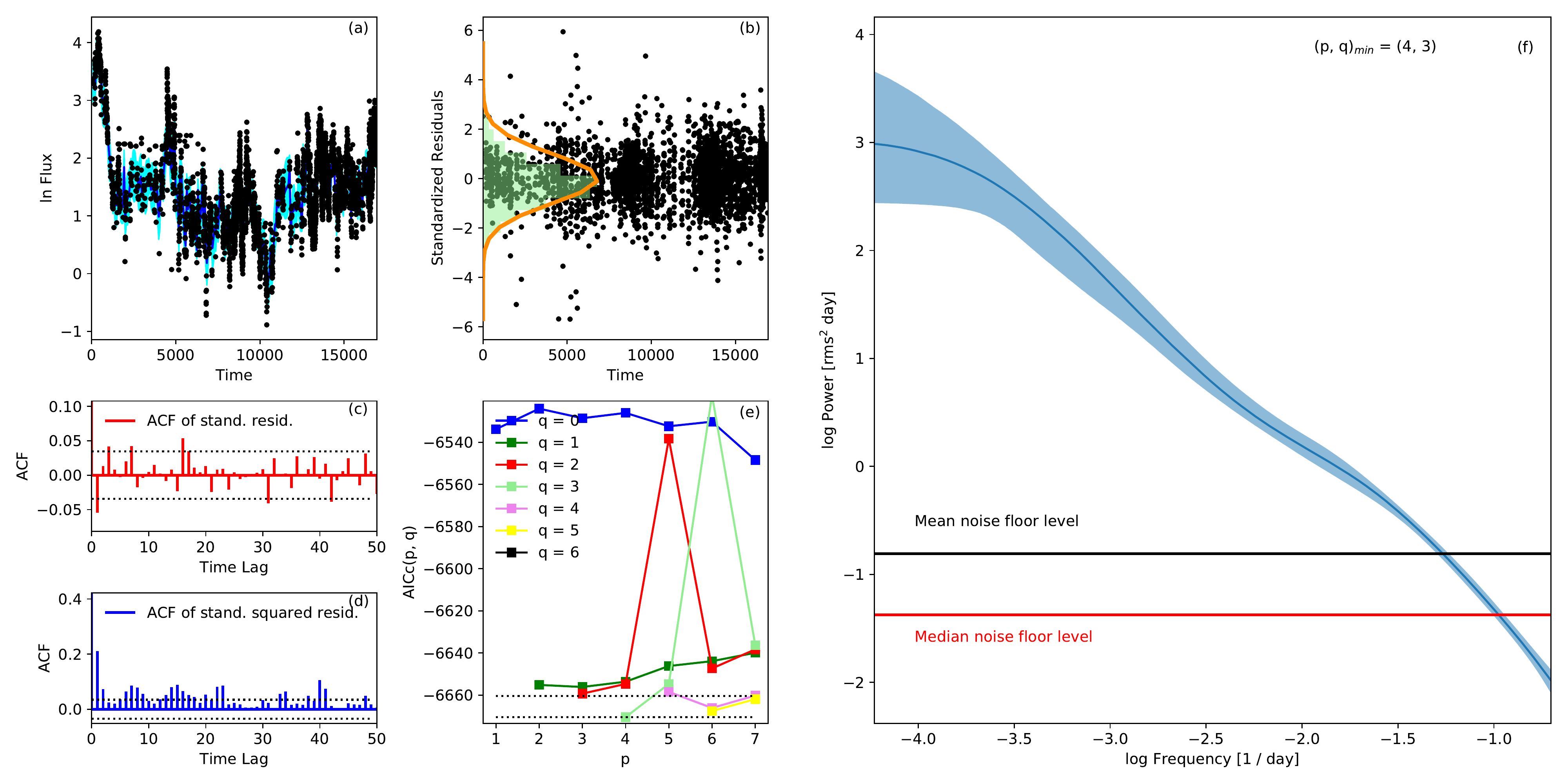}
\end{center}
\caption{As in Figure~\ref{fig:5}, only for the historical 1970--2017 optical data (optical(trun.); Table~\ref{tab:2}), also best fit by the CARMA(4,3) model.} 
\label{fig:8}
\end{figure*}

\begin{figure*}
\begin{center}
\includegraphics[width=1.0\textwidth]{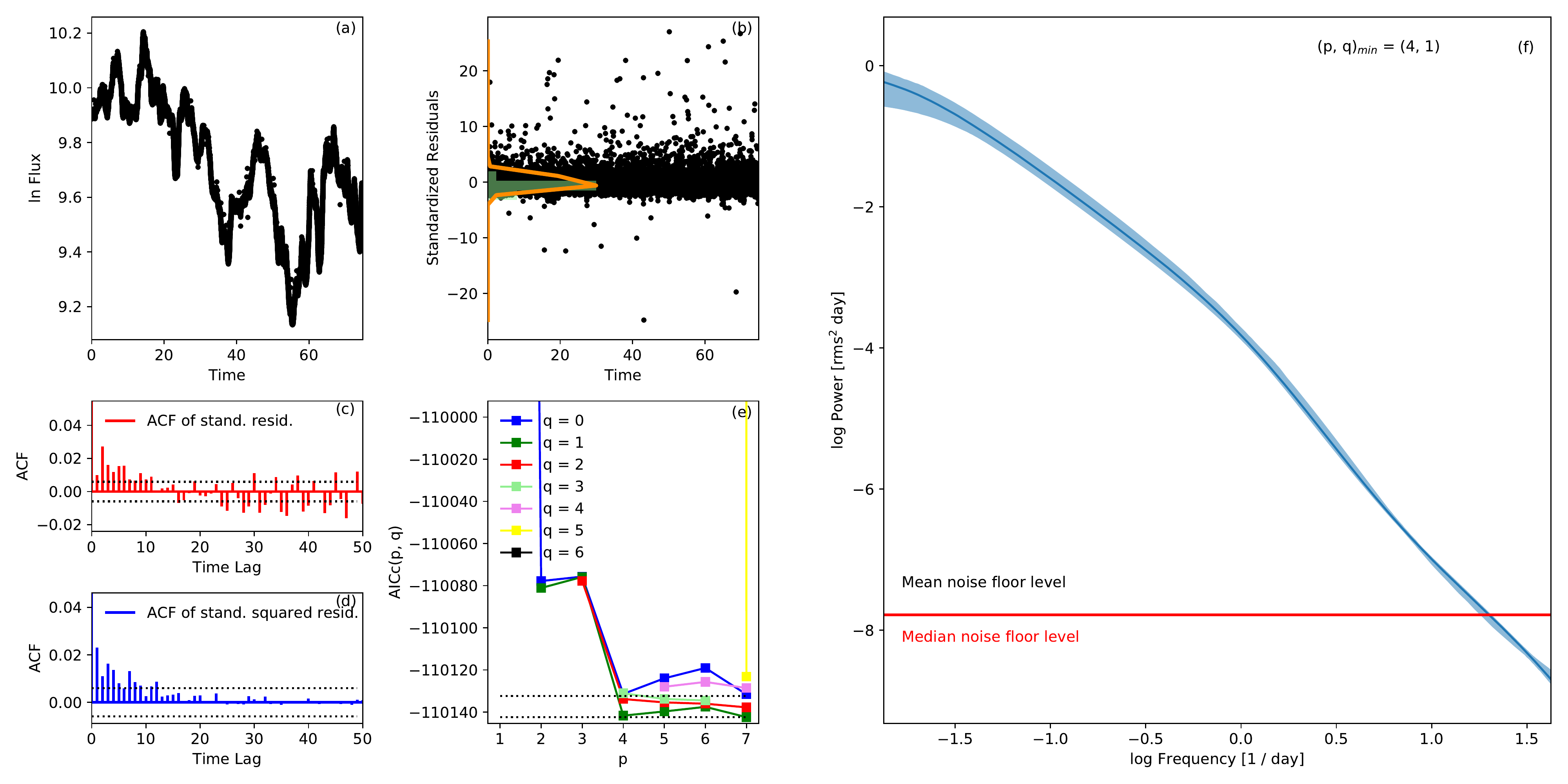}
\end{center}
\caption{As in Figure~\ref{fig:5}, but for the {\it Kepler} data, very well fit by the CARMA(4,1) model.}
\label{fig:9}
\end{figure*}

\begin{figure*}
\begin{center}
\includegraphics[width=1.0\textwidth]{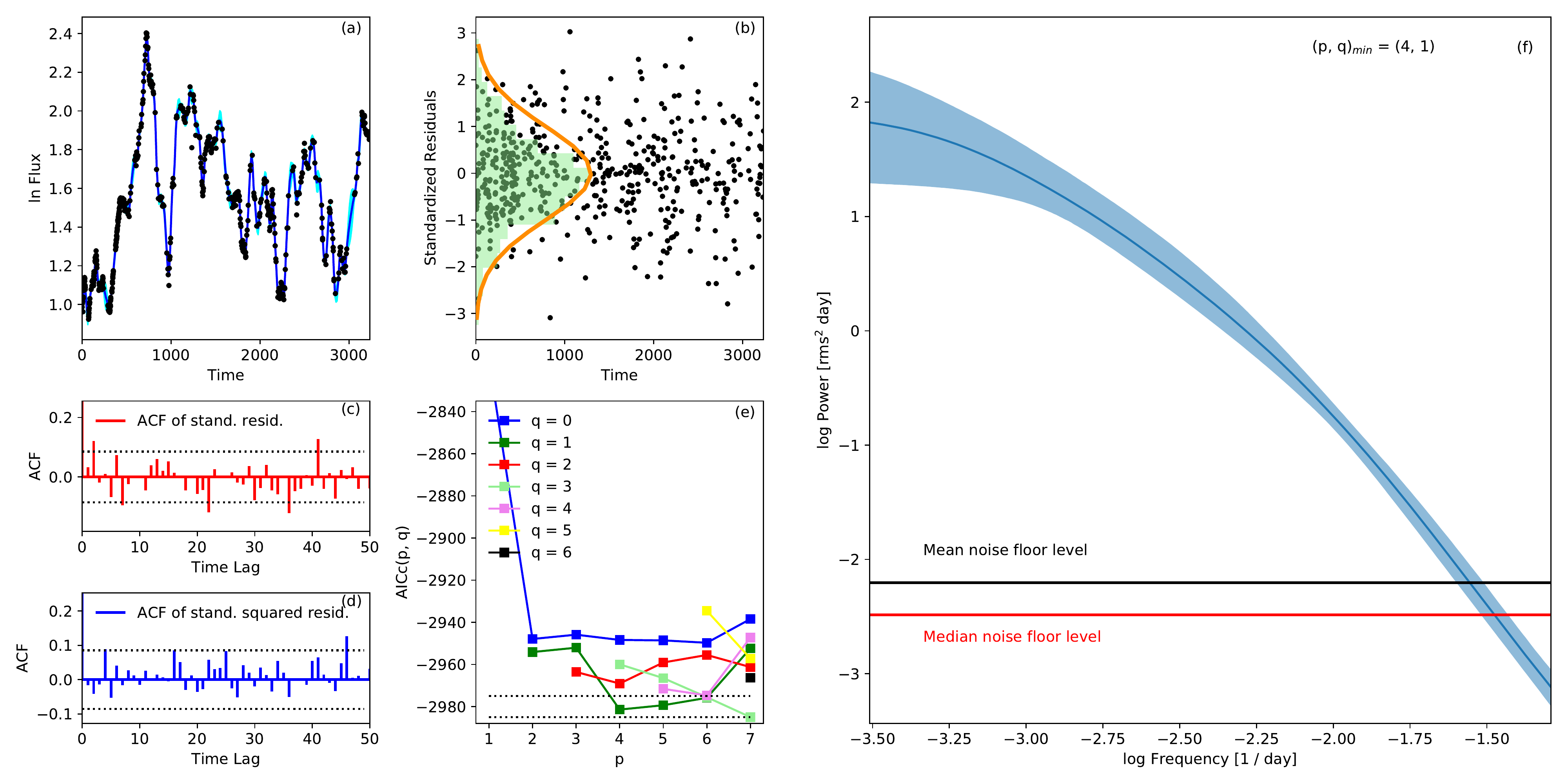}
\end{center}
\caption{As in Figure~\ref{fig:5}, but for the 15\,GHz OVRO data, best fit by the CARMA(4,1) model.}
\label{fig:10}
\end{figure*}

\begin{figure*}
\begin{center}
\includegraphics[width=1.0\textwidth]{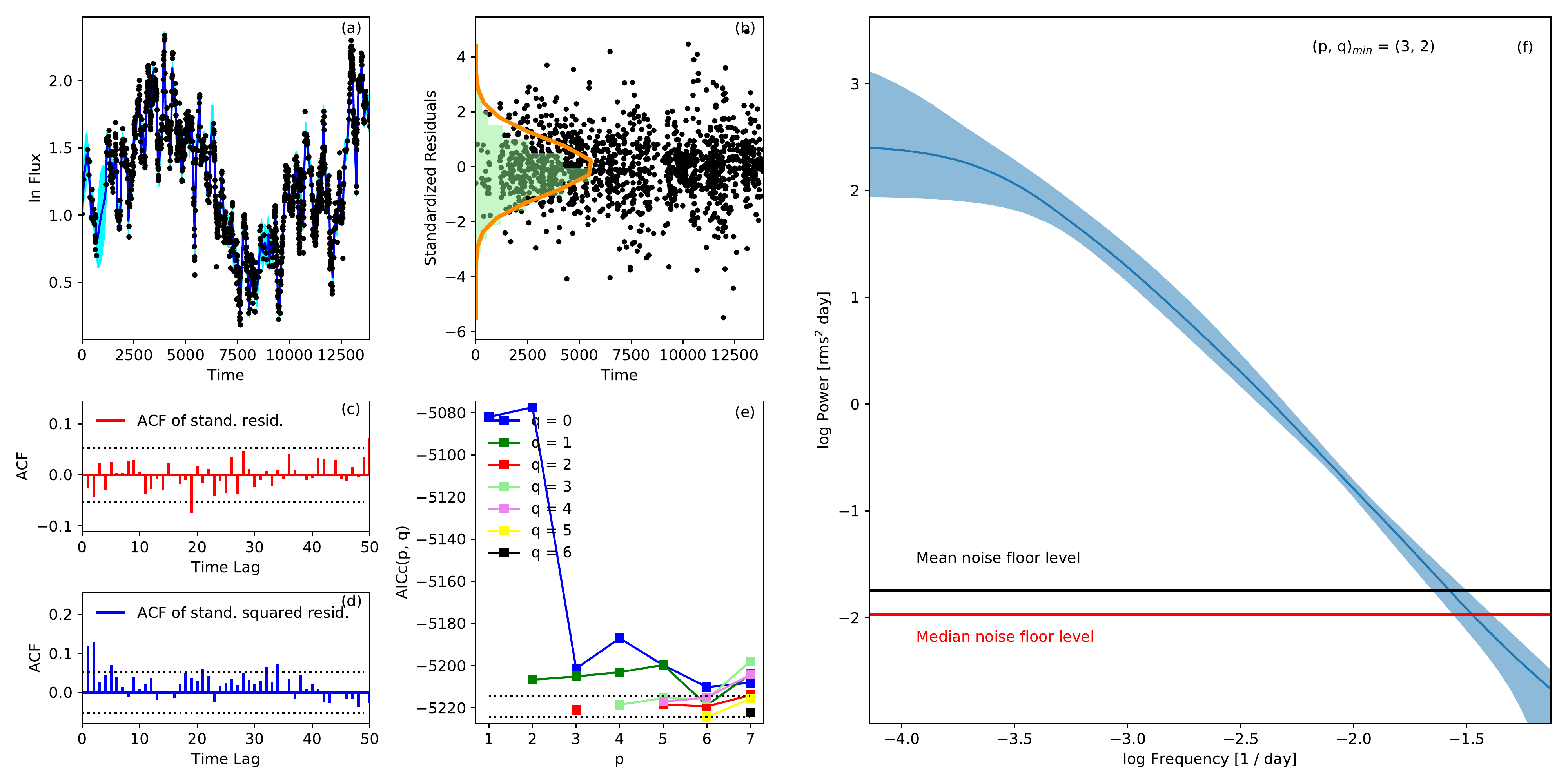}
\end{center}
\caption{As in Figure~\ref{fig:5}, but for the 14.5\,GHz UMRAO data, very well fit by the CARMA(3,2) model.}
\label{fig:11}
\end{figure*}

\begin{figure*}
\begin{center}
\includegraphics[width=1.0\textwidth]{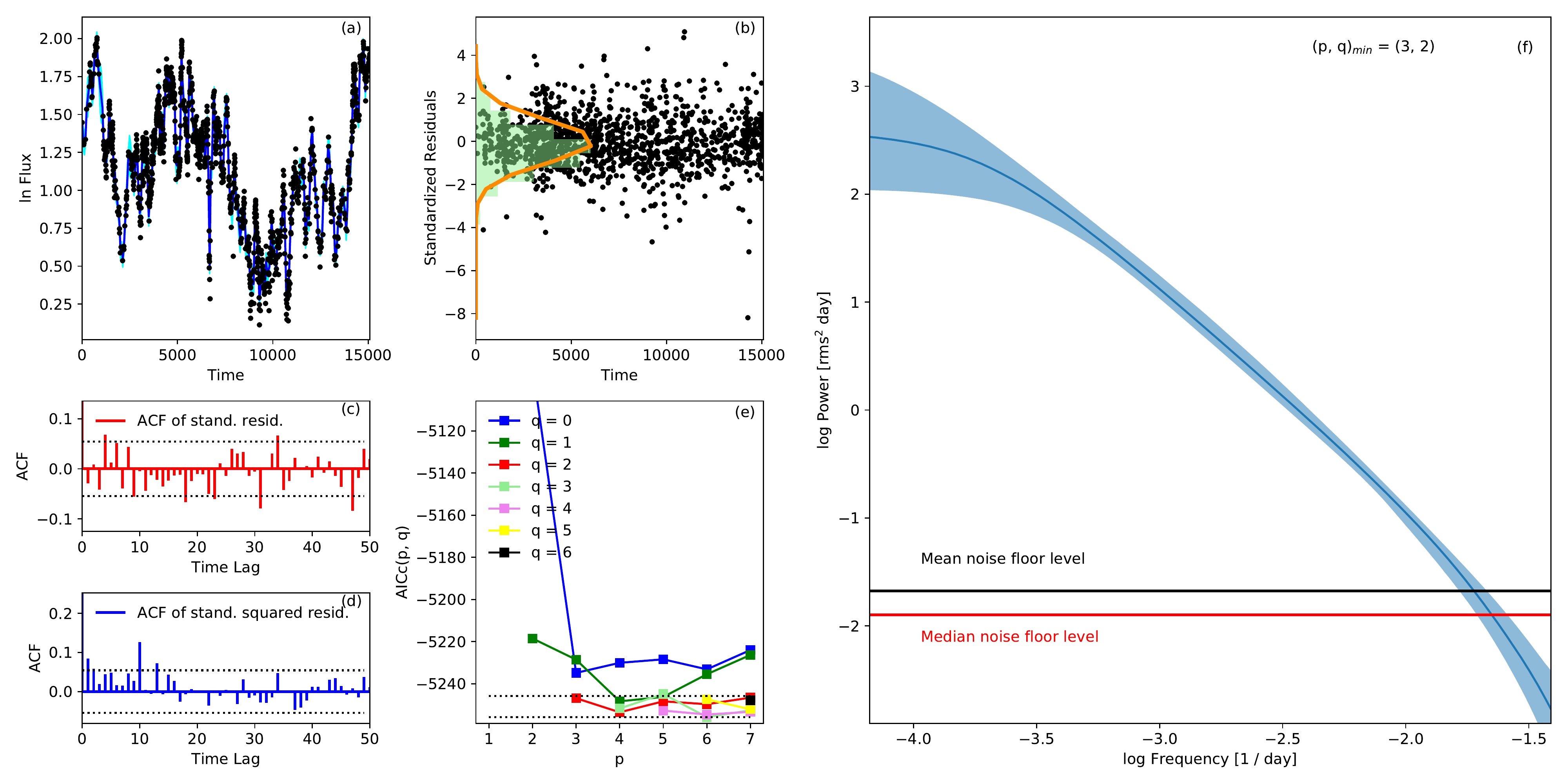}
\end{center}
\caption{As in Figure~\ref{fig:5}, but for the 8\,GHz UMRAO data, very well fit by the CARMA(3,2) model.}
\label{fig:12}
\end{figure*}

\begin{figure*}[t!]
\begin{center}
\includegraphics[width=1.0\textwidth]{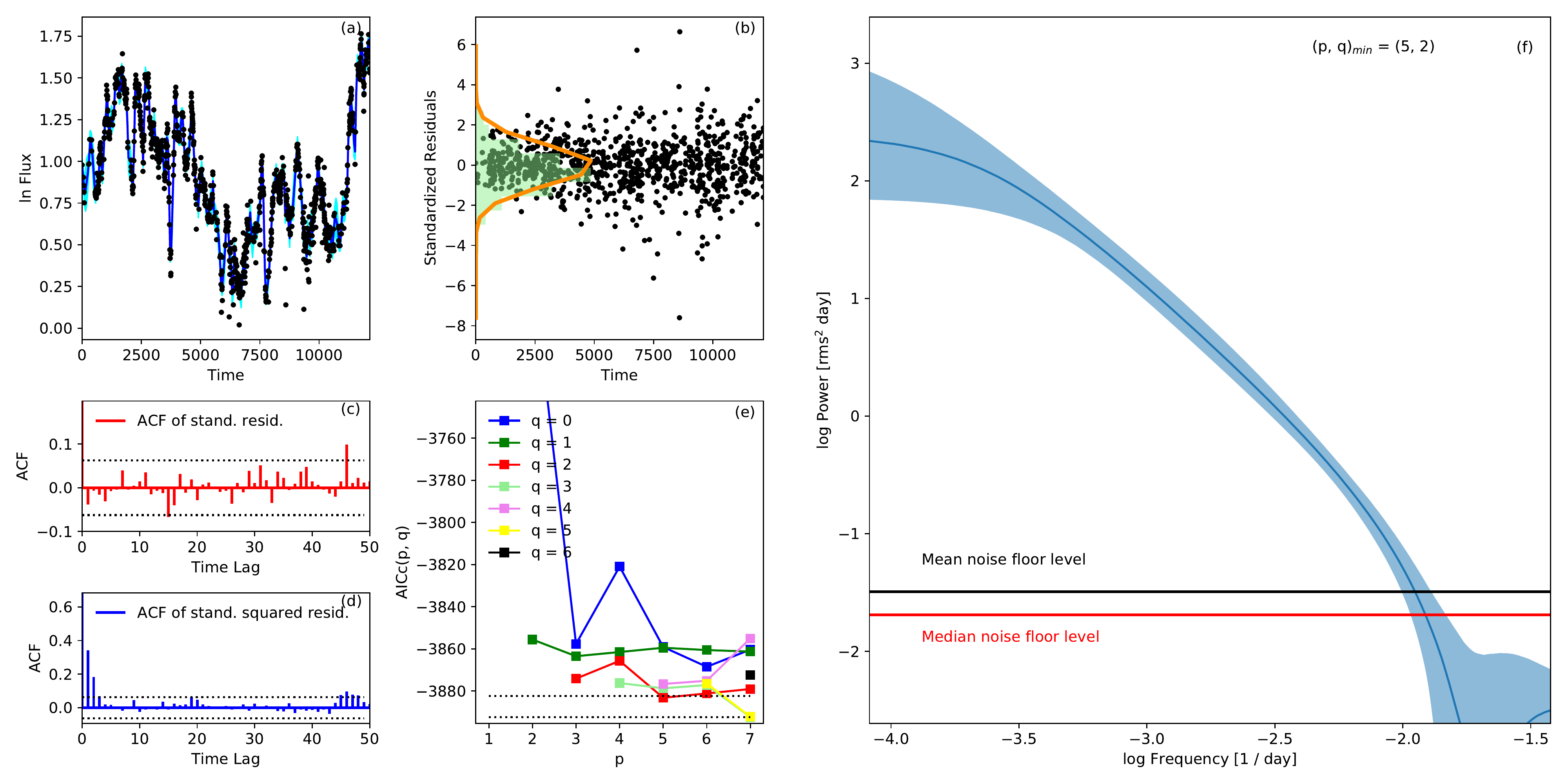}
\end{center}
\caption{As in Figure~\ref{fig:5}, but for the 4.8\,GHz UMRAO data, best fit by the CARMA(5,2) model.}
\label{fig:13}
\end{figure*}

\section{Results}
\label{sec:result}

The flux distributions of blazar light curves can be modeled nonlinearly, in the sense that they often can be represented as $y(t) = \exp[l(t)]$, where $l(t)$ is a linear Gaussian time series \citep[see, e.g.,][]{Edelson13,Kushwaha16,Hess17,Liodakis17}.  Hence, we have logarithmically transformed the light curves of OJ\,287 analyzed here (Figures~\ref{fig:1}--\ref{fig:3}), and then modeled them as Gaussian CARMA$(p,q)$ processes. For each light curve, the minimum $(p,q)$ order was selected by minimizing the AICc values on the grid $p=1,..,7$ and  $q=0,..,p-1$. For such, we ran the MCMC sampler for $\mathcal{N}$ iterations with the first $\mathcal{N}/2$ iterations discarded as a burn-in. Next, we employed the \citet{Gelman92} method as a diagnostic to analyze the chain convergence using the multiple-chain approach, allowing us to compare the ``within chain'' and ``between-chain'' variances. The number of iterations was chosen to derive the potential scale reduction factor for all the model parameters to be $< 1.001$. We select as the best-fit model the one produced by the pair of $(p,q)$ values having the lowest order within the range in which models are statistically indistinguishable from each other (i.e., minimum AICc $< 10$; see Section~\ref{sec:carma}). Figure~\ref{fig:15} in Appendix~\ref{app:B} shows, for comparison, the power spectra obtained for different $(p,q)$ parameters consistent with the null hypothesis of the CARMA process for the analyzed {\it Fermi-}LAT light curve (see Figure~\ref{fig:3}(a)).  

The results of the CARMA model fitting are presented in Figures~\ref{fig:5} (high-energy $\gamma$-rays), \ref{fig:6} (X-rays), \ref{fig:7}--\ref{fig:9} (optical), and \ref{fig:10}--\ref{fig:13} (radio). We plot the measured time series along with the modeled values based on the best-fit CARMA process (panels a), the standardized residuals and their distribution compared with the expected normal distributions (panels b), the corresponding ACFs compared with the $95\%$ confidence regions for a white-noise process (panels c), the squared ACFs compared with the $95\%$ confidence regions for a white-noise process (panels d), the AICc values for different $(p,q)$ pairs (panels e), and the resulting PSDs with $2\sigma$ confidence regions, as well as noise floor levels $P_{\rm stat}$ marked by horizontal lines (panels f). As shown, all the analyzed light curves are well represented by Gaussian CARMA processes, as the residuals from the model fitting follow the expected normal distributions with the ACFs and the squared ACFs lying within $2\sigma$ intervals for most of the temporal lags.} Note that because  some of the light curves are sparsely sampled (in particular, the historical optical and the {\it Swift}-XRT light curves), we estimated the noise floor level (Eq.~\ref{noise_stat}) with either ``mean'' or ``median'' sampling intervals.

Figure~\ref{fig:7} presents the PSDs for the entire available long-term (1900--2017) monitoring optical dataset, with typical daily sampling (hereafter ``optical(all)''). The majority of the optical data obtained before 1970 ($\sim 7$\% of the data points) have measurement uncertainties of order 20\%, owing to large calibration errors resulting from observations recorded on photographic plates (see \citealt{Hudec13} for a discussion of error estimation). The overall noise floor level in the derived long-term PSD is relatively high, firstly because of larger measurement uncertainties, and secondly because the mean sampling interval is large, $>12$ days. Therefore, we also derived the long-term PSD for the data obtained using only the good-quality photomultiplier tubes and CCD photometric measurements for the period 1970--2017, with typical measurement uncertainties $\sim 2$--5\% and typical sampling of 5 days (hereafter ``optical(trun.)''), as shown in Figure~\ref{fig:8}. Finally, the PSD corresponding to the continuous 72\,day-long monitoring {\it Kepler} data, with sampling down to $\sim 1$\,min (see Table~\ref{sec:obs}), is presented in Figure~\ref{fig:9}. The optical PSD of the blazar obtained by combining the PSDs generated with optical(all), optical(trun.), and the {\it Kepler} data covers an unprecedented frequency range of nearly 6 dex (from 117\,yr down to hour-long timescales), \emph{without} any gaps. The lower-frequency segment of the {\it Kepler} PSD is found to be consistent with a simple extrapolation of the optical PSD following from the long-term monitoring of the source.

Figure~\ref{fig:14} presents the composite multiwavelength PSDs of OJ\,287, truncated below the median noise floor level for each dataset. As seen, the PSDs derived using GHz-band radio light curves follow each other closely, on timescales ranging from decades to weeks/months. Also, there is a remarkable similarity between the radio and optical PSDs on variability timescales longer than $\sim 1$\,yr. On the other hand, on timescales shorter than one year, the X-ray and radio power spectra seem to resemble each other, with the variability amplitudes smaller than those observed in the optical. Importantly, the overall shape of the $\gamma$-ray PSD is clearly distinct from the (generally speaking, colored-noise type) PSDs derived at lower-frequency bands, at least over timescales probed by the length of the LAT light curve, down to the median LAT sampling interval. In particular, in the high-energy $\gamma$-ray range we see rather uncorrelated flux changes on timescales longer than $\simeq 157^{+350}_{-91}$\,d, manifesting as a flat (``white-noise'') segment in the derived variability power spectrum. We note in this context that \emph{some} low-frequency flattening can be noticed in basically all the PSDs modeled with CARMA; this is due to the fact that, by definition (Eq.~\ref{eq:carma}), the model applied includes the uncorrelated Gaussian (``driving'') term $\epsilon(t)$. However, only in the case of the $\gamma$-ray power spectrum could such a flattening be considered significant (note the $2\sigma$ confidence regions marked as blue areas in Figures~\ref{fig:5}--\ref{fig:13}, especially in the X-ray power spectrum shown in Figure~\ref{fig:13}).

\begin{figure*}[t!]
\begin{center}
\includegraphics[width=1.0\textwidth]{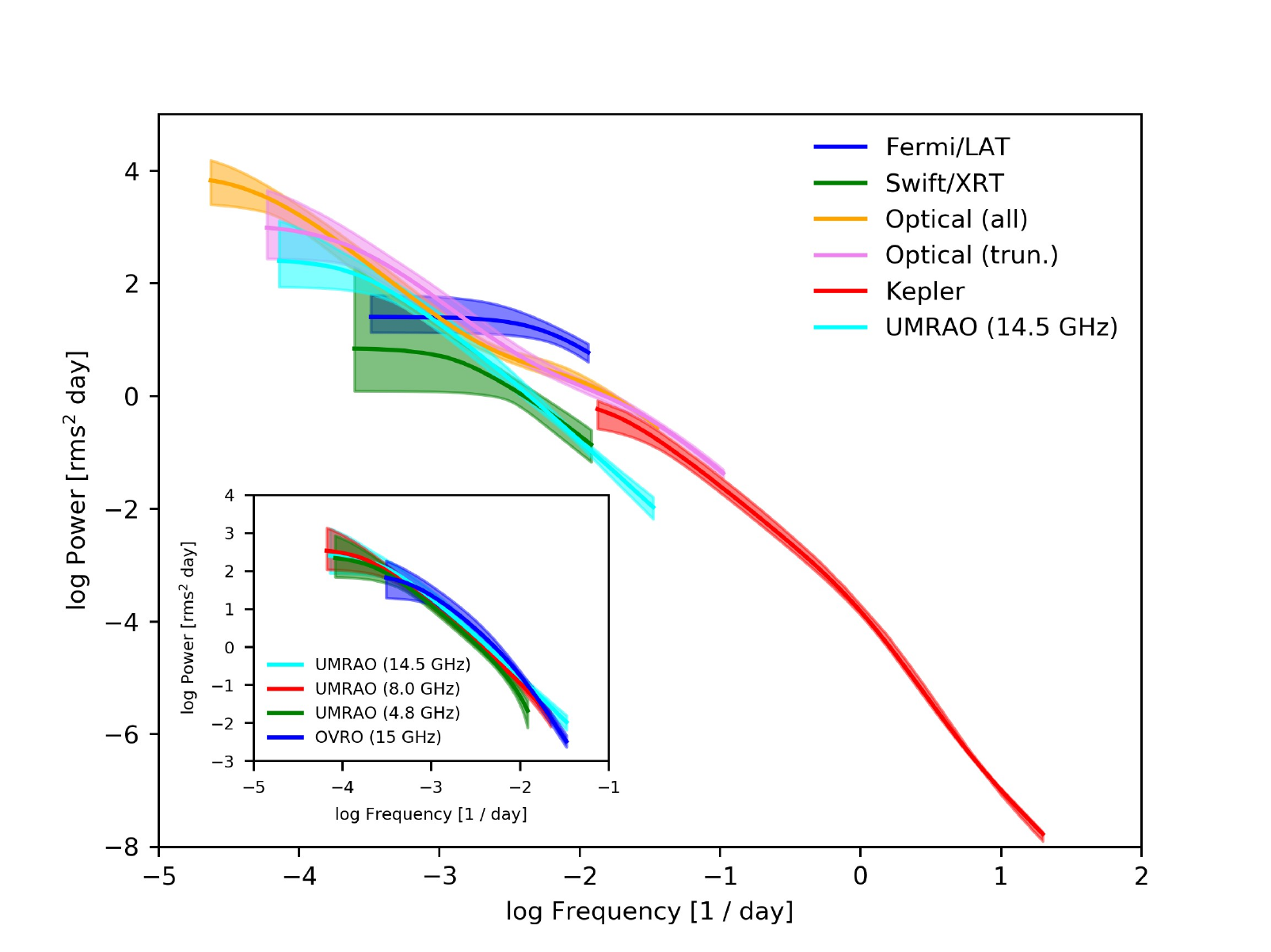}
\end{center}
\caption{Composite multiwavelength power spectra of OJ\,287, with the colored curves as described in the legend;  $2\sigma$ confidence intervals are shown by the corresponding shaded regions.}
\label{fig:14}
\end{figure*}

\section{Discussion and Conclusions}
\label{sec:conclusion}

The main findings from our analysis of the multiwavelength and multi-epoch measurements of OJ\,287 can be summarized as follows.
 
\begin{itemize}

\item[(i)] On timescales ranging from tens of years down to months, the power spectra derived at radio frequencies based on single-dish observations indicate a colored-noise character of the source variability.

\item[(ii)] Owing to the inclusion of the {\it Kepler} data, we were able to construct \emph{for the first time} the optical variability power spectrum of a blazar without any gaps across $\sim6$ dex in temporal frequencies. The modeled power spectrum is well represented by a higher-order CARMA process, on timescales ranging from 117\,yr down to $< 1$\,hr, but with no significant narrow features that could be identified with QPOs (see the next section below).

\item[(iii)] The power spectrum at X-ray photon energies, based on relatively sparsely sampled {\it Swift}-XRT data, could be characterized essentially only on timescales shorter than a year (down to months), where it resembles the radio power spectrum of the target. 

\item[(iv)] The high-energy $\gamma$-ray power spectrum of OJ\,287, modeled using the {\it Fermi}-LAT data, is noticeably different from the radio, optical, and X-ray power spectra of the source. In particular, in the framework of the CARMA modeling, this power spectrum is consistent with the CAR(1) process, according to the minimum AICc criterion adopted in this study, with an uncorrelated Gaussian (white) noise above the relaxation timescale of $\sim 150$\,d, and a correlated (colored) noise on timescales shorter than 150\,d.

\end{itemize}
 
\subsection{Periodicity in the long-term optical and {\it Fermi}-LAT light curves}

OJ\, 287 has become famous to a large extent because of the claims of $\sim 12$\,yr periodicity in its optical light curve \citep[see, e.g.,][for a recent review]{Valtonen16}. However, the present analysis does not reveal any well-defined peak in the power spectrum corresponding to this timescale (see Figure~\ref{fig:7}). On the other hand, even though the total duration of the optical light curve analyzed in this paper is $\sim 117$\,yr, the data obtained before 1970 are highly irregularly sampled \citep[see also][]{Hudec13}. The better sampled 1970--2017 light curve (see Figure~\ref{fig:8}) covers only $\sim 3$ of the claimed cycles, and as such is not sufficiently long to reveal any significant periodicity (a {\it sinusoidal component) over the colored-noise ({\it stochastic} component) of the} power spectrum \citep[see the discussion in][and Appendices~\ref{app:C} and ~\ref{app:D} here]{Vaughan16}. 

We also do not see any QPOs in either {\it Fermi-LAT} or optical data around 400 days, which were reported by \citet{2016AJ....151...54S} and \citet{2016ApJ...832...47B}, based on the LSP analysis \citep{Scargle82}. However, in those studies, the significance level for the detection of narrow spectral features in the periodograms was assigned based on a large number of mock light curves generated for the best-fit PSD model of the underlying colored-noise (stochastic) component \citep[e.g.,][]{Uttley02}. As emphasized by \citet[][]{Vaughan16}, a proper characterization of this stochastic component --- in particular, its exact slope (``color'') --- is crucial in this respect. On the other hand, if the periodicity is real, then the significance of the corresponding spectral feature in the periodogram should increase with the data length, as any periodic component will continue to oscillate around the same frequency while the stochastic component will smooth out. However, this inference is invalid if the observed periodicity is transitory in nature \citep[see the analysis and the discussion by][]{2016ApJ...832...47B}. Thus, the fact that we do not observe the reported periodicities in our longer datasets may simply be attributed to the transitory nature of the quasi-periodic signal. 

\subsection{Characteristic variability timescales}

As noted above, detailed CARMA modeling of the long-term optical monitoring data, together with the {\it Kepler} data, indicates a rather complex shape of the source power spectrum, with a more rapid decrease in the variability amplitudes with variability frequencies on timescales shorter than a day, when compared with the decrease observed at longer variability timescales. Interestingly, a fairly similar feature on broadly analogous timescales, manifesting as a break in a periodogram modelled as a power law, has been reported by \citet{2015ApJ...798...27I} in the X-ray power spectrum of the blazar Mrk\,421, based on a comparison between the {\it MAXI} and {\it ASCA} satellite data \citep[see also][]{2001ApJ...560..659K}. This break may indicate either a nonstationarity of the variability process in the source on timescales of order days and shorter, or --- if persistent --- it may signal some characteristic variability timescale in the system (in particular, the timescale below which there is a rapid decline in the variability power, although overall the variability process is still of a colored-noise type). Interestingly, the peak of the synchrotron component (in the spectral energy distribution representation) falls within the optical range for OJ\,287 (hence classified as a ``low-frequency-peaked'' BL Lac), and in the X-ray range for Mrk\,421 (a ``high-frequency-peaked'' BL Lac object). 

Turning to the {\it Fermi}-LAT light curve of OJ\,287, our CARMA modeling shows a clear break in the variability power spectrum: on the timescales longer than $\sim 150$\,d we see uncorrelated (white) noise, and on shorter timescales there is correlated (colored) noise. The 150\,d could therefore be identified with a characteristic relaxation timescale in the system, which is however related only to the production of high-energy $\gamma$-rays. Analogous breaks have frequently been reported in the optical and X-ray power spectra of radio-quiet AGN, on timescales of hundreds of days, depending on the black hole mass and the accretion rate in the studied systems \citep{Mchardy06,Kelly09,Kelly11}. In radio-quiet AGN, the observed optical and X-ray emission originate within the accretion disk, and hundreds-day timescales could be reconciled with the thermal timescales of the innermost parts of the disks. In blazar sources, on the other hand, the observed $\gamma$-ray fluxes are instead due to relativistic jets, and there is no obvious physical reason for a $\sim 150$\,d relaxation, unless one assumes a strong, almost one-to-one, coupling between the disk and the jet $\gamma$-ray variabilities \citep[but see][]{AG17,2017ApJ...843...81O}. Interestingly, a similar feature of the high-energy $\gamma$-ray power spectra, breaking from white to colored noise, has been reported before by  \citet{2014ApJ...786..143S} in the {\it Fermi}-LAT light curves of the BL Lac objects PKS\,2155$-$304 and 3C\,66A, albeit on shorter timescales of about a month, and in the long-term optical monitoring data for PKS\,2155$-$304 by \citet{2011A&A...531A.123K} on a timescale of  $\sim 1000$\,days.

\subsection{Multiwavelength power spectra}

In our recent analysis of the multiwavelength power spectra of the low-frequency-peaked BL Lac object PKS\,0735+178 \citep{AG17}, which however did not include any X-ray data, and was moreover based on the discrete Fourier transform method (with linear interpolation), we found that the statistical character of the $\gamma$-ray flux changes is different from that of the radio and optical flux changes. Specifically, the high-energy $\gamma$-ray power spectrum of the source was found to be consistent with a flickering noise, while the radio and optical power spectra with a pure red noise. There we suggested that this finding could be understood in terms of a model where the blazar synchrotron variability is generated by the underlying single stochastic process, and the inverse-Compton variability by a linear superposition of such processes, within a highly nonuniform portion of the outflow extending from the jet base up to the $\lesssim 1$\,pc-scale distances. 

The more robust analysis of the much higher quality multiwavelength data for OJ\,287 presented in this paper, based on the CARMA modeling, is to a large extent consistent with the findings reported before for PKS\,0735+178 by \citet{AG17}. That is, the overall slope of the high-energy $\gamma$-ray PSD in OJ\,287 is significantly flatter than the slopes of radio or optical PSDs, and also the colored-noise-type variability at optical frequencies occurs over a very broad range of variability timescales, from decades to hours. However, the new finding emerging from the analysis presented here is that (i) there may be a more rapid decrease of variability amplitudes with variability frequency in the optical power spectrum of OJ\,287 on timescales shorter than a day, (ii) the X-ray power spectrum of the source resembles the radio power spectra on timescales ranging from a year down to months/weeks, and that (iii) the high-energy $\gamma$-ray power spectrum of the blazar reveals a  relaxation timescale on the order of 150\,d (which is not seen in the power spectra at lower photon energies). 

The interpretation of the above novel findings is not straightforward, keeping in mind that, in the particular case of OJ\,287, the observed X-ray emission seems to be mostly produced by the inverse-Compton process involving the lowest-energy electrons, being only occasionally dominated by the high-energy tail of the synchrotron continuum \citep[e.g.,][]{2009PASJ...61.1011S}. A possible resolution may lie in the scenario where the high-energy $\gamma$-ray emission does not constitute the high-energy tail of the broad-band inverse-Compton continuum extending from X-ray photon energies, but instead is caused by a distinct (spectrally and spatially) electron population, peaked at the highest electron energies, and distributed rather exclusively within the innermost parts of the jet, thus being much more responsive to the faster modulations associated with the accretion-disk events as compared to the outer parts of the jet.

\begin{acknowledgements} 

We thank the anonymous referees for constructive comments on the manuscript. The authors thank also B.~C. Kelly and J. Ballet for useful discussions during the revision of the paper.

A.G. and M.O. acknowledge support from the Polish National Science Centre (NCN) through the grant 2012/04/A/ST9/00083. D.K.W. and A.G. acknowledge the support from 2013/09/B/ST9/00026. {\L}.S. and V.M. are supported by Polish NSC grant UMO-2016/22/E/ST9/00061. M.S. acknowledges the support of 2012/07/B/ST9/04404. S.Z. acknowledges the support of 2013/09/B/ST9/00599 and 2017/27/B/ST9/01855. R.H acknowledges GA CR grant 13-33324S. A.S. and M.So. were supported by National Aeronautics and Space Administration (NASA) contract NAS8-03060 (Chandra X-ray Center). M.So. also acknowledges Polish NCN grant OPUS 2014/13/B/ST9/00570. A.V.F. is grateful for support from National Science Foundation (NSF) grant AST-1211916, NASA grant NNX12AF12G, the TABASGO Foundation, the Christopher R. Redlich Fund, and the Miller Institute for Basic Research in Science (U.C. Berkeley). Research at Lick Observatory is partially supported by a generous gift from Google.

This research has made use of data from the University of Michigan Radio Astronomy Observatory, which has been supported by the University of Michigan and by a series of grants from the NSF, most recently AST-0607523, and NASA Fermi grants NNX09AU16G, NNX10AP16G, and NNX11AO13G. The OVRO 40-m Telescope Fermi Blazar Monitoring Program is supported by NASA under awards NNX08AW31G and NNX11A043G, and by the NSF under awards AST-0808050 and AST-1109911. Based on observations obtained with telescopes of the University Observatory Jena, which is operated by the Astrophysical Institute of the Friedrich-Schiller University. 

The \textit{Fermi} LAT Collaboration acknowledges generous ongoing support from a number of agencies and institutes that have supported both the development and the operation of the LAT as well as scientific data analysis. These include NASA and the Department of Energy (DOE) in the United States, the Commissariat \`a l'Energie Atomique and the Centre National de la Recherche Scientifique / Institut National de Physique Nucl\'eaire et de Physique des Particules in France, the Agenzia Spaziale Italiana and the Istituto Nazionale di Fisica Nucleare in Italy, the Ministry of Education, Culture, Sports, Science and Technology (MEXT), High Energy Accelerator Research Organization (KEK), and Japan Aerospace Exploration Agency (JAXA) in Japan, and the K.~A.~Wallenberg Foundation, the Swedish Research Council, and the Swedish National Space Board in Sweden. Additional support for science analysis during the operations phase is gratefully acknowledged from the Istituto Nazionale di Astrofisica in Italy and the Centre National d'\'Etudes Spatiales in France. This work was performed in part under DOE Contract DE-AC02-76SF00515.

\end{acknowledgements} 

\appendix

\section{Characterization of power spectral density with unevenly sampled data series }
\label{app:A}

In our test simulations an artificial light curve is generated using the method of \citet{Emmanoulopoulos13}, assuming a pure red-noise ($a=2$) power spectrum. The data points are evenly sampled with a sampling interval of one day, and the total duration of the simulated time series is 1,000 days. We then kept 30\% of the data selected at random times to mimic an unevenly sampled dataset. The simulated light curves are presented in Figure \ref{fig:15}(a). Figure~\ref{fig:15}(b) shows the corresponding power spectra derived using the LSP method and the FT of the ACF, while Figure \ref{fig:15}(c) shows the response of the spectral window for the FT of the ACF. We refer the reader to \citet{Scargle82} and \citet{Edelson88} or the Appendix of \citet{AG17} for the mathematical details of these methods. The derived power spectrum is fitted with simple power-law form $P (\nu_k) \propto \nu_k^{-a}$ where $a$ is the slope of the power spectrum over the frequencies corresponding to 1,000 days down to 2 days.  As shown, the derived PSDs (slopes $a=0.3$ and $-0.1$ for the LSP and the FT of the ACF, respectively), are very different from the true PSD ($a = 2$). The flattening of the derived power spectra in the high variability frequency range in the case of the LSP is due to the DOF problem, which introduces  artificial power in the high-frequency range of the spectrum. Note that this is not caused by the addition of the Gaussian noise (white noise; $a=0$), as the simulated light curve is free of such an effect. In the case of the FT of the ACF, which is free of such a ``missing data points'' problem, the flattening is instead caused by a substantial amount of variability power provided by the spectral window function. For FT of the ACF method, the resulting power spectrum is convolved with the spectral window function, leading to flatter power-law slopes. In principal, one can de-convolve the power spectrum knowing the spectral window function (Figure 15(c)). As shown, the spectral window function becomes more and more noisy towards higher frequencies and therefore the de-convolved power spectrum will be more and more uncertain towards higher and higher frequencies.

\begin{figure*}[h!]
\hbox{
\hspace*{0.1cm}\includegraphics[height=6.0cm,width=8.5cm]{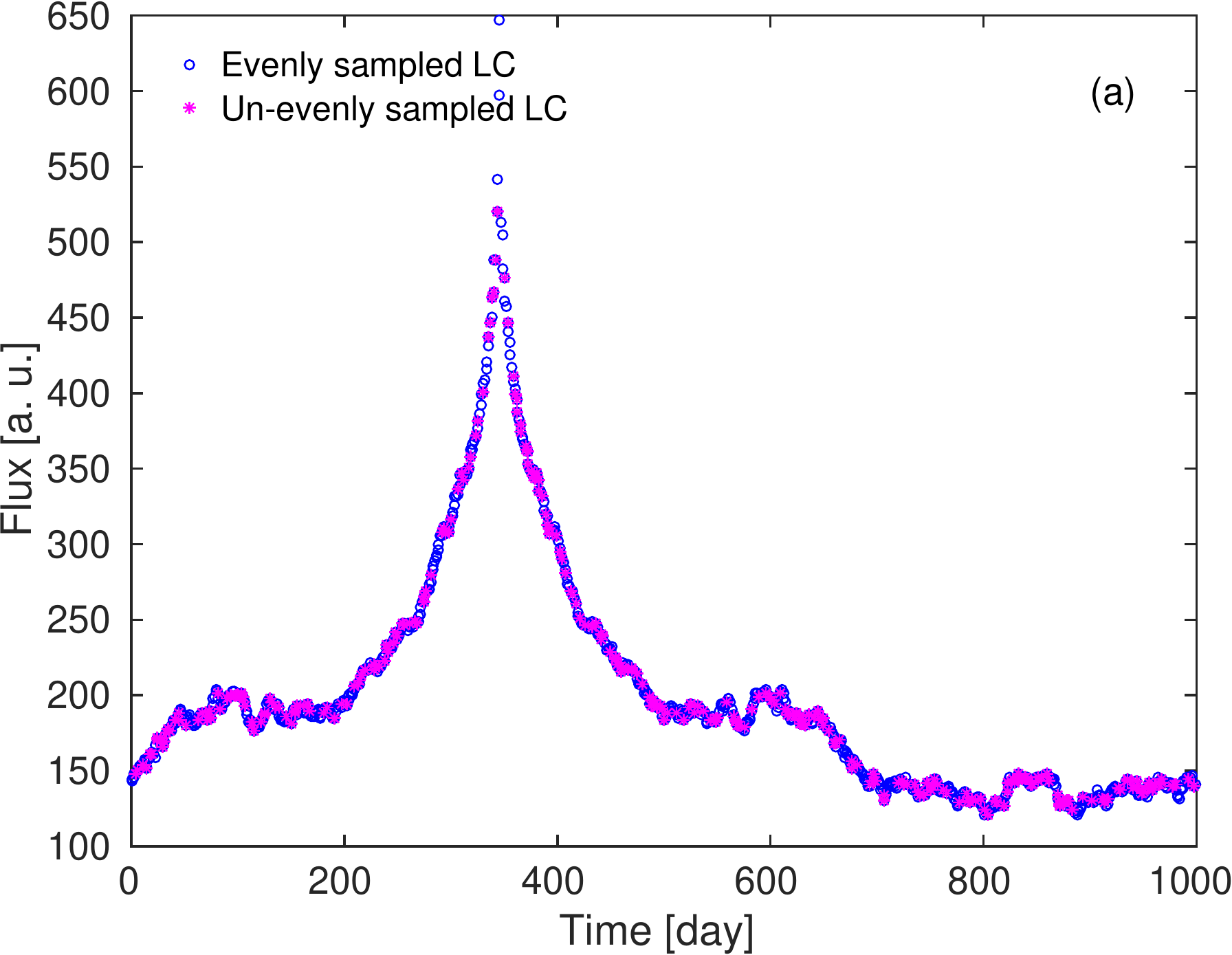}
\hspace*{0.1cm}\includegraphics[height=6.0cm,width=8.5cm]{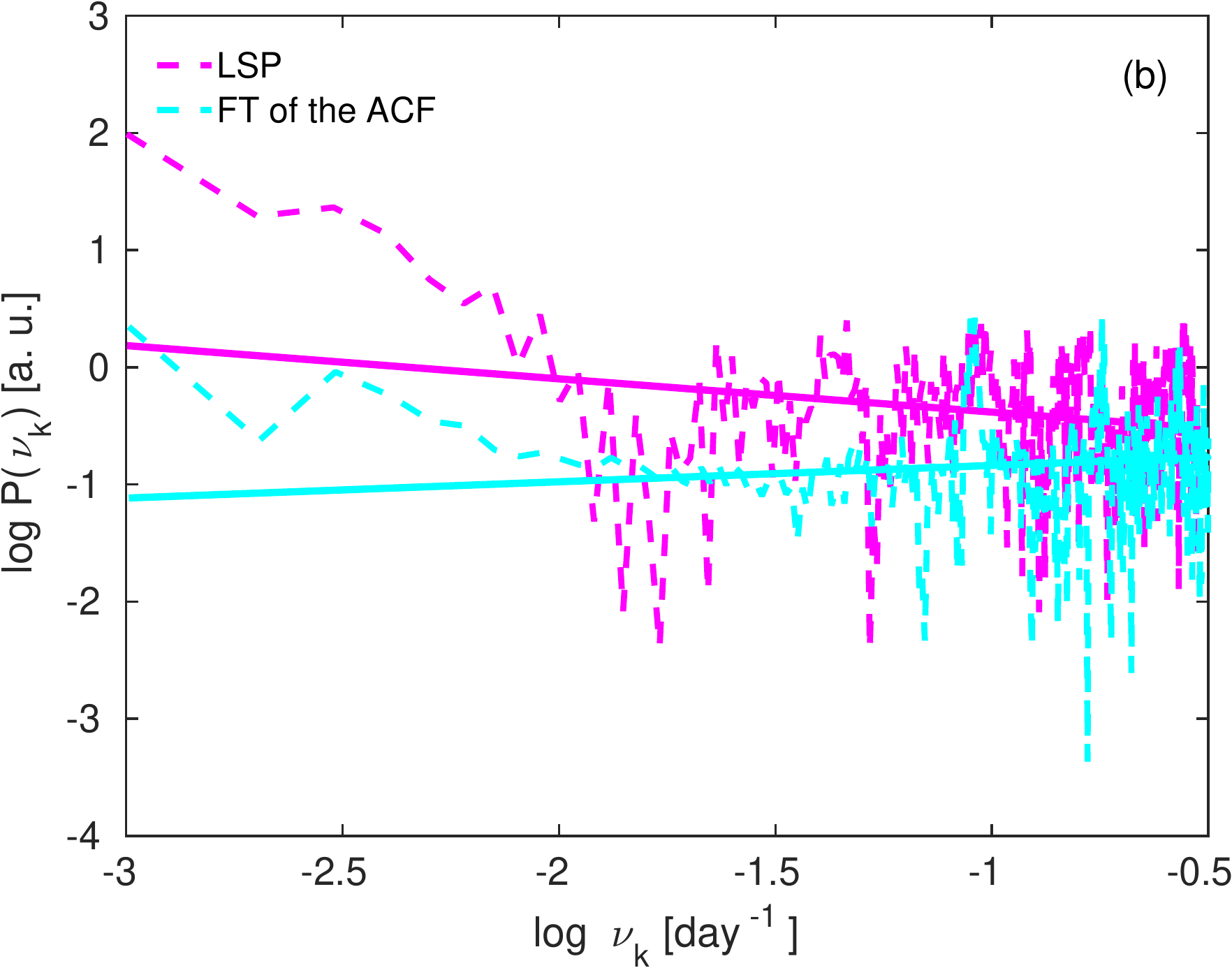}
}
\hspace*{4.0cm}\includegraphics[height=6.0cm,width=8.5cm]{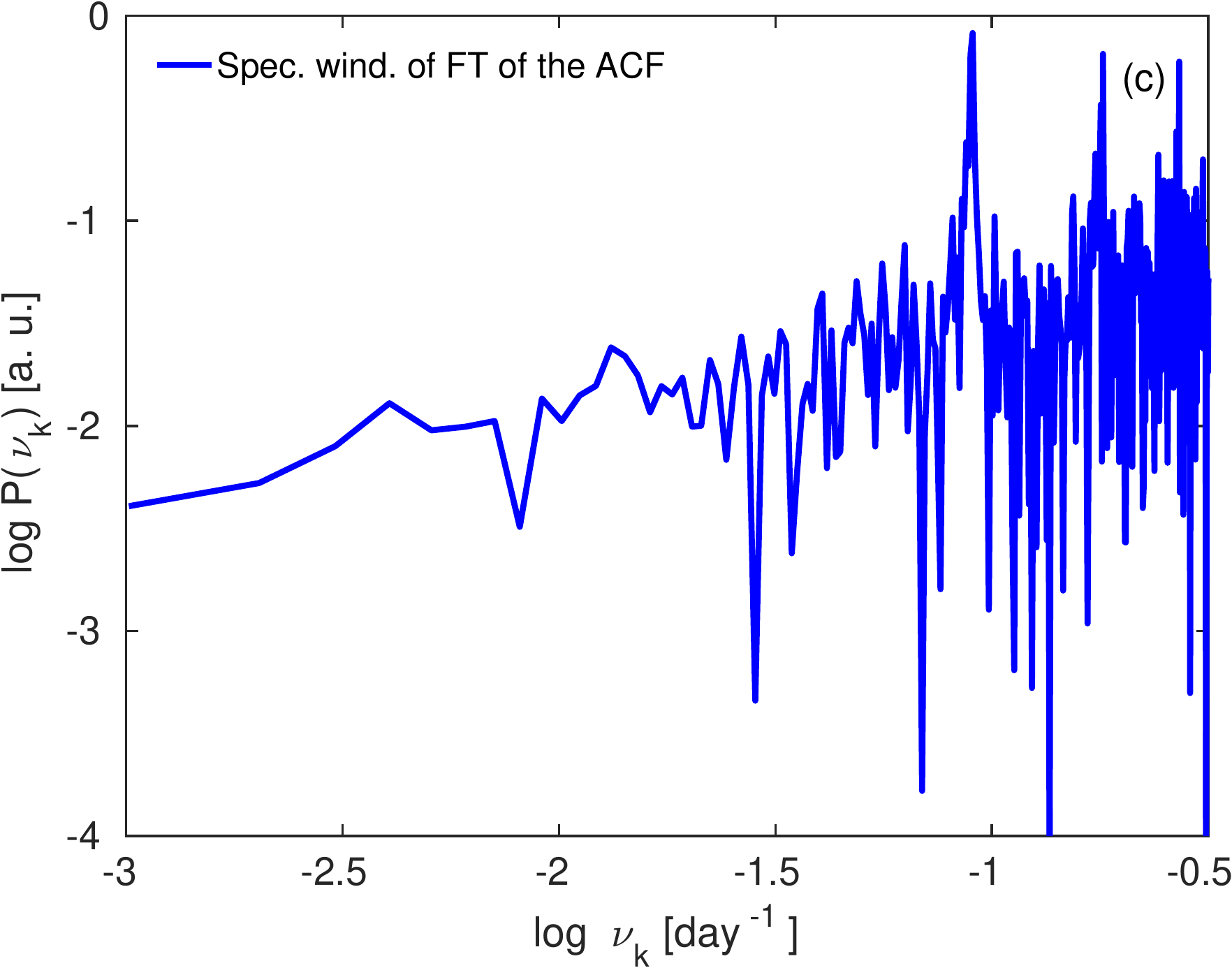}
\caption{(a) Simulated red-noise light curve ($a = 2$), containing 1,000 data points with a sampling interval of one day (blue open circles), along with 30\% of the time series selected at random times to mimic an unevenly sampled dataset (magenta stars). (b) The corresponding power spectra of the unevenly sampled light curve derived using the LSP method (resulting in $a=0.3$; magenta dashed curve), and the FT of the ACF (returning $a = -0.1$; cyan dashed curve). (c) Spectral window of power spectrum estimated from FT of the ACF (see Appendix~\ref{app:A} for details).}
\label{fig:15}
\end{figure*}

\section{Selection of $(p,q)$ parameters for the CARMA process}
\label{app:B}

The order of the CARMA$(p,q)$ process is chosen based on how close the model is to the data using the minimum AICc criterion adopted in the present study. It has been argued that the models for various pairs of $(p,q)$ values for which the minimum AICc is within 10 are not statistically indistinguishable from each other. Figure \ref{fig:16} shows power spectra corresponding to a few sets of $(p,q)$ parameters of the analyzed {\it Fermi}-LAT light curve. As these overlap substantially, we choose the lowest-order model --- CARMA(1,0) --- as the best-fitting model to describe the high-energy $\gamma-$ray variability in OJ\,287.
\begin{figure*}[h!]
\begin{center}
\includegraphics[width=0.8\textwidth]{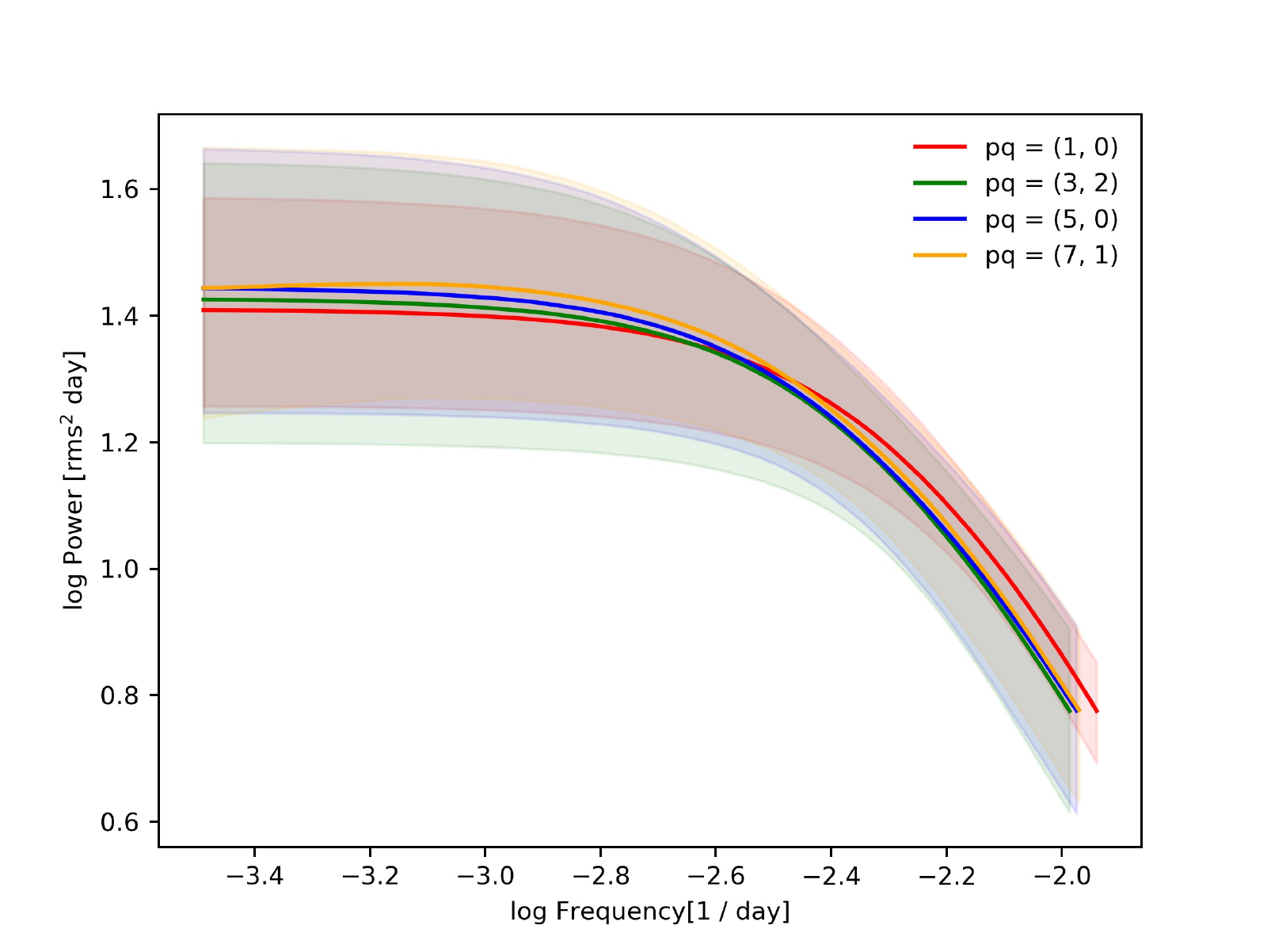}
\end{center}
\caption{Power spectra corresponding to a few sets of $(p,q)$ parameters of the analyzed {\it Fermi}-LAT light curve of OJ\,287. The corresponding $1\sigma$ confidence intervals are denoted by shaded areas. See Appendix~\ref{app:B} for details.}
\label{fig:16}
\end{figure*}

\section{Lack of $\sim 12$\,yr QPO in the optical light curve}
\label{app:C}

Here we discuss in more detail the lack of the claimed $\sim12$\,yr quasi-periodicity in the analyzed optical light curve. To demonstrate the robustness of the CARMA modeling in detecting QPO features against the background of red noise in the power spectrum for a finite time series covering only a few periods, an artificial light curve is generated with three components: (1) a pure red-noise ($\beta= 2$) power spectrum using the method of \citet{Emmanoulopoulos13}, (2) a sinusoidal component with a 12\,yr period, and (3) a Gaussian white noise with mean 0 and standard deviation 0.1 representing measurement uncertainty. The data points are evenly sampled with a sampling interval of one day, and the total duration of the simulated time series is 14,600 days (40\,yr, corresponding to our relatively better sampled optical light curve starting from 1970; see Table~\ref{tab:2}). Next, we kept 20\% of the data selected at random times to mimic an unevenly sampled dataset. Figure~\ref{fig:17}(a--e) presents the results of CARMA modeling on our simulated light curve, while Figure~\ref{fig:17}(f) shows the computed power spectrum. Since the simulated light curve covers only $\sim3$ periods, we do not detect a clear peak on the $\sim 12$\,yr timescale against the background of a red-noise power spectrum and Gaussian white noise, in accordance with the discussion by \citet{Vaughan16}.   

\begin{figure*}[h!]
\begin{center}
\hbox{
\includegraphics[width=\textwidth]{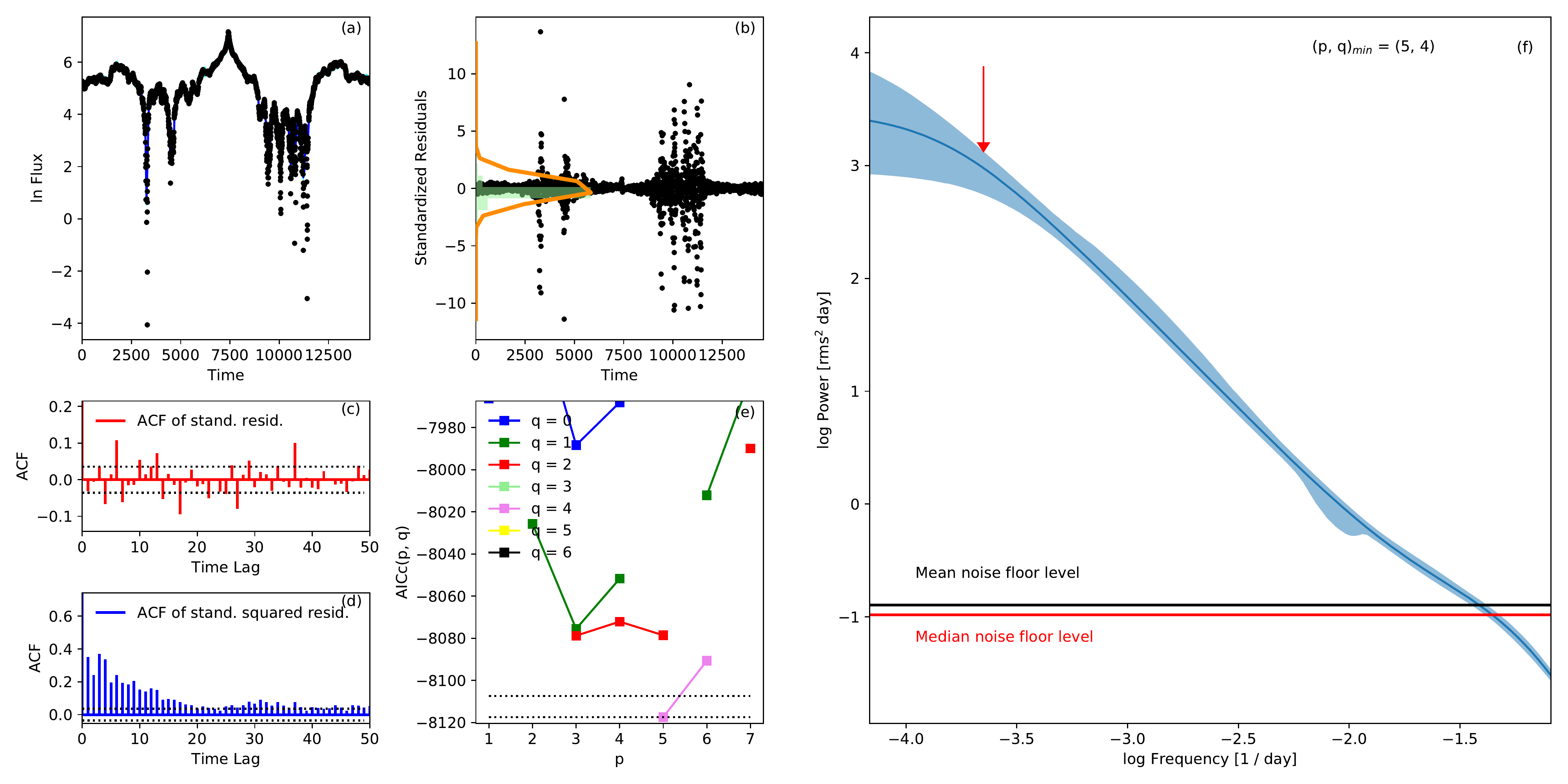}
}
\end{center}
\caption{Results of the CARMA modeling of the simulated red-noise ($\beta=2$) light curve with 12\,yr periodicity, as described in Appendix~\ref{app:C}; the best-fit model is $(p=3, q=2)$. The red arrow marks the position of the putative QPO. }
\label{fig:17}
\end{figure*}

\section{Comparison of power spectra from the historical optical light curve}
\label{app:D}

Using the historical optical(all) light curve (1900--2017) analysed here, we calculated the power spectrum using the DFT method \citep[see ][for mathematical details on computing PSD using the DFT method]{AG17}. Figure~\ref{fig:18} shows the resulting power spectrum obtained using CARMA modeling and the DFT method. Within a 3$\sigma$ confidence interval, the two methods give compatible results. A mild peak $\sim 12$\,yr from the DFT method is within the 3$\sigma$ confidence intervals estimated from the CARMA modeling.    

\begin{figure*}[h!]
\begin{center}
\includegraphics[width=0.8\textwidth]{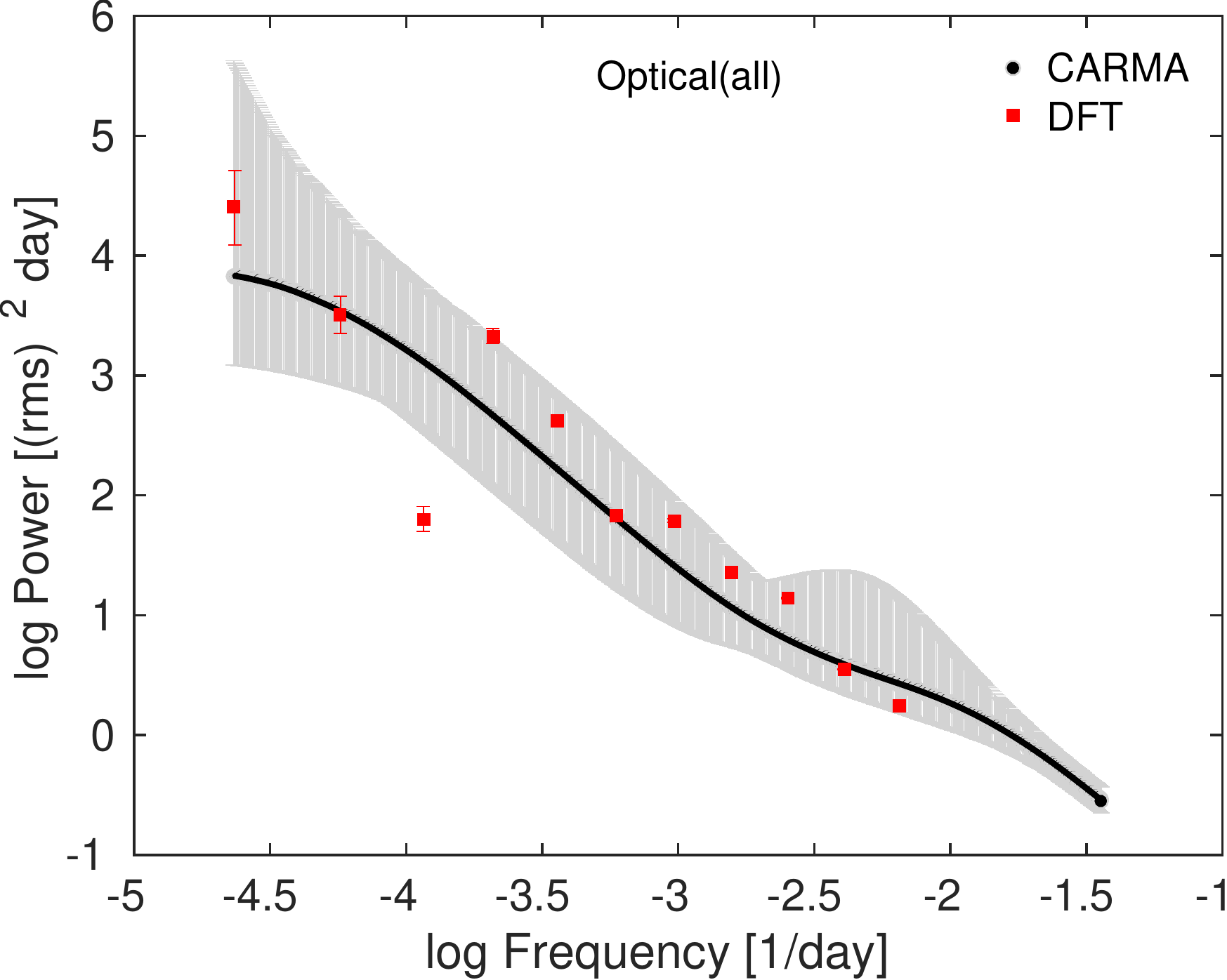}
\end{center}
\caption{The optical power spectrum of OJ\,287 computed using the DFT method (red) and CARMA modeling (black, with gray shaded regions corresponding to $3\sigma$ confidence intervals), as discussed in Appendix~\ref{app:D}.}
\label{fig:18}
\end{figure*}


\clearpage

\begin{thebibliography}{}

\bibitem[{{Abdalla} {et~al.}(2017){Abdalla}, {Abramowski}, {Aharonian}, {Ait
  Benkhali}, {Akhperjanian}, {Andersson}, {Ang{\"u}ner}, {Arrieta}, {Aubert},
  \& et~al.}]{Hess17}
{Abdalla}, H., {Abramowski}, A., {Aharonian}, F., {et~al.} 2017, \aap, 598, A39

\bibitem[{{Abdo} {et~al.}(2009){Abdo}, {Ackermann}, {Ajello}, {Atwood},
  {Axelsson}, {Baldini}, {Ballet}, {Band}, {Barbiellini}, {Bastieri}, \&
  et~al.}]{2009ApJS..183...46A}
{Abdo}, A.~A., {Ackermann}, M., {Ajello}, M., {et~al.} 2009, \apjs, 183, 46

\bibitem[{{Abdo} {et~al.}(2010){Abdo}, {Ackermann}, {Ajello}, {Allafort},
  {Antolini}, {Atwood}, {Axelsson}, {Baldini}, {Ballet}, {Barbiellini},
  {Bastieri}, {Baughman}, {Bechtol}, {Bellazzini}, {Berenji}, {Blandford},
  {Bloom}, {Bogart}, {Bonamente}, {Borgland}, {Bouvier}, {Bregeon}, {Brez},
  {Brigida}, {Bruel}, {Buehler}, {Burnett}, {Buson}, {Caliandro}, {Cameron},
  {Cannon}, {Caraveo}, {Carrigan}, {Casandjian}, {Cavazzuti}, {Cecchi}, {{\c
  C}elik}, {Celotti}, {Charles}, {Chekhtman}, {Chen}, {Cheung}, {Chiang},
  {Ciprini}, {Claus}, {Cohen-Tanugi}, {Conrad}, {Costamante}, {Cotter},
  {Cutini}, {D'Elia}, {Dermer}, {de Angelis}, {de Palma}, {De Rosa}, {Digel},
  {Silva}, {Drell}, {Dubois}, {Dumora}, {Escande}, {Farnier}, {Favuzzi},
  {Fegan}, {Ferrara}, {Focke}, {Fortin}, {Frailis}, {Fukazawa}, {Funk},
  {Fusco}, {Gargano}, {Gasparrini}, {Gehrels}, {Germani}, {Giebels},
  {Giglietto}, {Giommi}, {Giordano}, {Giroletti}, {Glanzman}, {Godfrey},
  {Grandi}, {Grenier}, {Grondin}, {Grove}, {Guiriec}, {Hadasch}, {Harding},
  {Hayashida}, {Hays}, {Healey}, {Hill}, {Horan}, {Hughes}, {Iafrate}, {Itoh},
  {J{\'o}hannesson}, {Johnson}, {Johnson}, {Johnson}, {Johnson}, {Kamae},
  {Katagiri}, {Kataoka}, {Kawai}, {Kerr}, {Kn{\"o}dlseder}, {Kuss}, {Lande},
  {Latronico}, {Lavalley}, {Lemoine-Goumard}, {Llena Garde}, {Longo},
  {Loparco}, {Lott}, {Lovellette}, {Lubrano}, {Madejski}, {Makeev}, {Malaguti},
  {Massaro}, {Mazziotta}, {McConville}, {McEnery}, {McGlynn}, {Michelson},
  {Mitthumsiri}, {Mizuno}, {Moiseev}, {Monte}, {Monzani}, {Morselli},
  {Moskalenko}, {Murgia}, {Nolan}, {Norris}, {Nuss}, {Ohno}, {Ohsugi},
  {Omodei}, {Orlando}, {Ormes}, {Ozaki}, {Paneque}, {Panetta}, {Parent},
  {Pelassa}, {Pepe}, {Pesce-Rollins}, {Piranomonte}, {Piron}, {Porter},
  {Rain{\`o}}, {Rando}, {Razzano}, {Reimer}, {Reimer}, {Reposeur}, {Ripken},
  {Ritz}, {Rodriguez}, {Romani}, {Roth}, {Ryde}, {Sadrozinski}, {Sanchez},
  {Sander}, {Saz Parkinson}, {Scargle}, {Sgr{\`o}}, {Shaw}, {Siskind}, {Smith},
  {Spandre}, {Spinelli}, {Starck}, {Stawarz}, {Strickman}, {Suson}, {Tajima},
  {Takahashi}, {Takahashi}, {Tanaka}, {Taylor}, {Thayer}, {Thayer}, {Thompson},
  {Tibaldo}, {Torres}, {Tosti}, {Tramacere}, {Ubertini}, {Uchiyama}, {Usher},
  {Vasileiou}, {Vilchez}, {Villata}, {Vitale}, {Waite}, {Wallace}, {Wang},
  {Winer}, {Wood}, {Yang}, {Ylinen}, \& {Ziegler}}]{2010ApJ...715..429Abdo}
---. 2010, \apj, 715, 429

\bibitem[{{Abdollahi} {et~al.}(2017){Abdollahi}, {Ackermann}, {Ajello},
  {Albert}, {Baldini}, {Ballet}, {Barbiellini}, {Bastieri}, {Becerra Gonzalez},
  {Bellazzini}, {Bissaldi}, {Blandford}, {Bloom}, {Bonino}, {Bottacini},
  {Bregeon}, {Bruel}, {Buehler}, {Buson}, {Cameron}, {Caragiulo}, {Caraveo},
  {Cavazzuti}, {Cecchi}, {Chekhtman}, {Cheung}, {Chiaro}, {Ciprini}, {Conrad},
  {Costantin}, {Costanza}, {Cutini}, {D'Ammando}, {de Palma}, {Desai},
  {Desiante}, {Digel}, {Di Lalla}, {Di Mauro}, {Di Venere}, {Donaggio},
  {Drell}, {Favuzzi}, {Fegan}, {Ferrara}, {Focke}, {Franckowiak}, {Fukazawa},
  {Funk}, {Fusco}, {Gargano}, {Gasparrini}, {Giglietto}, {Giomi}, {Giordano},
  {Giroletti}, {Glanzman}, {Green}, {Grenier}, {Grove}, {Guillemot}, {Guiriec},
  {Hays}, {Horan}, {Jogler}, {J{\'o}hannesson}, {Johnson}, {Kocevski}, {Kuss},
  {La Mura}, {Larsson}, {Latronico}, {Li}, {Longo}, {Loparco}, {Lovellette},
  {Lubrano}, {Magill}, {Maldera}, {Manfreda}, {Mayer}, {Mazziotta},
  {Michelson}, {Mitthumsiri}, {Mizuno}, {Monzani}, {Morselli}, {Moskalenko},
  {Negro}, {Nuss}, {Ohsugi}, {Omodei}, {Orienti}, {Orlando}, {Paliya},
  {Paneque}, {Perkins}, {Persic}, {Pesce-Rollins}, {Petrosian}, {Piron},
  {Porter}, {Principe}, {Rain{\`o}}, {Rando}, {Razzano}, {Razzaque}, {Reimer},
  {Reimer}, {Sgr{\`o}}, {Simone}, {Siskind}, {Spada}, {Spandre}, {Spinelli},
  {Stawarz}, {Suson}, {Takahashi}, {Tanaka}, {Thayer}, {Thompson}, {Torres},
  {Torresi}, {Tosti}, {Troja}, {Vianello}, \& {Wood}}]{Abdollahi17}
{Abdollahi}, S., {Ackermann}, M., {Ajello}, M., {et~al.} 2017, \apj, 846, 34

\bibitem[{{Acero} {et~al.}(2015){Acero}, {Ackermann}, {Ajello}, {Albert},
  {Atwood}, {Axelsson}, {Baldini}, {Ballet}, {Barbiellini}, {Bastieri},
  {Belfiore}, {Bellazzini}, {Bissaldi}, {Blandford}, {Bloom}, {Bogart},
  {Bonino}, {Bottacini}, {Bregeon}, {Britto}, {Bruel}, {Buehler}, {Burnett},
  {Buson}, {Caliandro}, {Cameron}, {Caputo}, {Caragiulo}, {Caraveo},
  {Casandjian}, {Cavazzuti}, {Charles}, {Chaves}, {Chekhtman}, {Cheung},
  {Chiang}, {Chiaro}, {Ciprini}, {Claus}, {Cohen-Tanugi}, {Cominsky}, {Conrad},
  {Cutini}, {D'Ammando}, {de Angelis}, {DeKlotz}, {de Palma}, {Desiante},
  {Digel}, {Di Venere}, {Drell}, {Dubois}, {Dumora}, {Favuzzi}, {Fegan},
  {Ferrara}, {Finke}, {Franckowiak}, {Fukazawa}, {Funk}, {Fusco}, {Gargano},
  {Gasparrini}, {Giebels}, {Giglietto}, {Giommi}, {Giordano}, {Giroletti},
  {Glanzman}, {Godfrey}, {Grenier}, {Grondin}, {Grove}, {Guillemot}, {Guiriec},
  {Hadasch}, {Harding}, {Hays}, {Hewitt}, {Hill}, {Horan}, {Iafrate}, {Jogler},
  {J{\'o}hannesson}, {Johnson}, {Johnson}, {Johnson}, {Johnson}, {Kamae},
  {Kataoka}, {Katsuta}, {Kuss}, {La Mura}, {Landriu}, {Larsson}, {Latronico},
  {Lemoine-Goumard}, {Li}, {Li}, {Longo}, {Loparco}, {Lott}, {Lovellette},
  {Lubrano}, {Madejski}, {Massaro}, {Mayer}, {Mazziotta}, {McEnery},
  {Michelson}, {Mirabal}, {Mizuno}, {Moiseev}, {Mongelli}, {Monzani},
  {Morselli}, {Moskalenko}, {Murgia}, {Nuss}, {Ohno}, {Ohsugi}, {Omodei},
  {Orienti}, {Orlando}, {Ormes}, {Paneque}, {Panetta}, {Perkins},
  {Pesce-Rollins}, {Piron}, {Pivato}, {Porter}, {Racusin}, {Rando}, {Razzano},
  {Razzaque}, {Reimer}, {Reimer}, {Reposeur}, {Rochester}, {Romani},
  {Salvetti}, {S{\'a}nchez-Conde}, {Saz Parkinson}, {Schulz}, {Siskind},
  {Smith}, {Spada}, {Spandre}, {Spinelli}, {Stephens}, {Strong}, {Suson},
  {Takahashi}, {Takahashi}, {Tanaka}, {Thayer}, {Thayer}, {Thompson},
  {Tibaldo}, {Tibolla}, {Torres}, {Torresi}, {Tosti}, {Troja}, {Van Klaveren},
  {Vianello}, {Winer}, {Wood}, {Wood}, {Zimmer}, \& {Fermi-LAT
  Collaboration}}]{Acero15}
{Acero}, F., {Ackermann}, M., {Ajello}, M., {et~al.} 2015, \apjs, 218, 23

\bibitem[{{Acero} {et~al.}(2016){Acero}, {Ackermann}, {Ajello}, {Albert},
  {Baldini}, {Ballet}, {Barbiellini}, {Bastieri}, {Bellazzini}, {Bissaldi},
  {Bloom}, {Bonino}, {Bottacini}, {Brandt}, {Bregeon}, {Bruel}, {Buehler},
  {Buson}, {Caliandro}, {Cameron}, {Caragiulo}, {Caraveo}, {Casandjian},
  {Cavazzuti}, {Cecchi}, {Charles}, {Chekhtman}, {Chiang}, {Chiaro}, {Ciprini},
  {Claus}, {Cohen-Tanugi}, {Conrad}, {Cuoco}, {Cutini}, {D'Ammando}, {de
  Angelis}, {de Palma}, {Desiante}, {Digel}, {Di Venere}, {Drell}, {Favuzzi},
  {Fegan}, {Ferrara}, {Focke}, {Franckowiak}, {Funk}, {Fusco}, {Gargano},
  {Gasparrini}, {Giglietto}, {Giordano}, {Giroletti}, {Glanzman}, {Godfrey},
  {Grenier}, {Guiriec}, {Hadasch}, {Harding}, {Hayashi}, {Hays}, {Hewitt},
  {Hill}, {Horan}, {Hou}, {Jogler}, {J{\'o}hannesson}, {Kamae}, {Kuss},
  {Landriu}, {Larsson}, {Latronico}, {Li}, {Li}, {Longo}, {Loparco},
  {Lovellette}, {Lubrano}, {Maldera}, {Malyshev}, {Manfreda}, {Martin},
  {Mayer}, {Mazziotta}, {McEnery}, {Michelson}, {Mirabal}, {Mizuno}, {Monzani},
  {Morselli}, {Nuss}, {Ohsugi}, {Omodei}, {Orienti}, {Orlando}, {Ormes},
  {Paneque}, {Pesce-Rollins}, {Piron}, {Pivato}, {Rain{\`o}}, {Rando},
  {Razzano}, {Razzaque}, {Reimer}, {Reimer}, {Remy}, {Renault},
  {S{\'a}nchez-Conde}, {Schaal}, {Schulz}, {Sgr{\`o}}, {Siskind}, {Spada},
  {Spandre}, {Spinelli}, {Strong}, {Suson}, {Tajima}, {Takahashi}, {Thayer},
  {Thompson}, {Tibaldo}, {Tinivella}, {Torres}, {Tosti}, {Troja}, {Vianello},
  {Werner}, {Wood}, {Wood}, {Zaharijas}, \& {Zimmer}}]{Acero16}
---. 2016, \apjs, 223, 26

\bibitem[{{Ackermann} {et~al.}(2016){Ackermann}, {Anantua}, {Asano}, {Baldini},
  {Barbiellini}, {Bastieri}, {Becerra Gonzalez}, {Bellazzini}, {Bissaldi},
  {Blandford}, {Bloom}, {Bonino}, {Bottacini}, {Bruel}, {Buehler}, {Caliandro},
  {Cameron}, {Caragiulo}, {Caraveo}, {Cavazzuti}, {Cecchi}, {Cheung}, {Chiang},
  {Chiaro}, {Ciprini}, {Cohen-Tanugi}, {Costanza}, {Cutini}, {DAmmando}, {de
  Palma}, {Desiante}, {Digel}, {Di Lalla}, {Di Mauro}, {Di Venere}, {Drell},
  {Favuzzi}, {Fegan}, {Ferrara}, {Fukazawa}, {Funk}, {Fusco}, {Gargano},
  {Gasparrini}, {Giglietto}, {Giordano}, {Giroletti}, {Grenier}, {Guillemot},
  {Guiriec}, {Hayashida}, {Hays}, {Horan}, {J{\'o}hannesson}, {Kensei},
  {Kocevski}, {Kuss}, {La Mura}, {Larsson}, {Latronico}, {Li}, {Longo},
  {Loparco}, {Lott}, {Lovellette}, {Lubrano}, {Madejski}, {Magill}, {Maldera},
  {Manfreda}, {Mayer}, {Mazziotta}, {Michelson}, {Mirabal}, {Mizuno},
  {Monzani}, {Morselli}, {Moskalenko}, {Nalewajko}, {Negro}, {Nuss}, {Ohsugi},
  {Orlando}, {Paneque}, {Perkins}, {Pesce-Rollins}, {Piron}, {Pivato},
  {Porter}, {Principe}, {Rando}, {Razzano}, {Razzaque}, {Reimer}, {Scargle},
  {Sgr{\`o}}, {Sikora}, {Simone}, {Siskind}, {Spada}, {Spinelli}, {Stawarz},
  {Thayer}, {Thompson}, {Torres}, {Troja}, {Uchiyama}, {Yuan}, \&
  {Zimmer}}]{2016ApJ...824L..20A}
{Ackermann}, M., {Anantua}, R., {Asano}, K., {et~al.} 2016, \apjl, 824, L20

\bibitem[{{Aharonian} {et~al.}(2007){Aharonian}, {Akhperjanian}, {Bazer-Bachi},
  {Behera}, {Beilicke}, {Benbow}, {Berge}, {Bernl{\"o}hr}, {Boisson}, {Bolz},
  {Borrel}, {Boutelier}, {Braun}, {Brion}, {Brown}, {B{\"u}hler},
  {B{\"u}sching}, {Bulik}, {Carrigan}, {Chadwick}, {Clapson}, {Chounet},
  {Coignet}, {Cornils}, {Costamante}, {Degrange}, {Dickinson},
  {Djannati-Ata{\"i}}, {Domainko}, {Drury}, {Dubus}, {Dyks}, {Egberts},
  {Emmanoulopoulos}, {Espigat}, {Farnier}, {Feinstein}, {Fiasson},
  {F{\"o}rster}, {Fontaine}, {Funk}, {Funk}, {F{\"u}{\ss}ling}, {Gallant},
  {Giebels}, {Glicenstein}, {Gl{\"u}ck}, {Goret}, {Hadjichristidis}, {Hauser},
  {Hauser}, {Heinzelmann}, {Henri}, {Hermann}, {Hinton}, {Hoffmann}, {Hofmann},
  {Holleran}, {Hoppe}, {Horns}, {Jacholkowska}, {de Jager}, {Kendziorra},
  {Kerschhaggl}, {Kh{\'e}lifi}, {Komin}, {Kosack}, {Lamanna}, {Latham}, {Le
  Gallou}, {Lemi{\`e}re}, {Lemoine-Goumard}, {Lenain}, {Lohse}, {Martin},
  {Martineau-Huynh}, {Marcowith}, {Masterson}, {Maurin}, {McComb}, {Moderski},
  {Moulin}, {de Naurois}, {Nedbal}, {Nolan}, {Olive}, {Orford}, {Osborne},
  {Ostrowski}, {Panter}, {Pedaletti}, {Pelletier}, {Petrucci}, {Pita},
  {P{\"u}hlhofer}, {Punch}, {Ranchon}, {Raubenheimer}, {Raue}, {Rayner},
  {Renaud}, {Ripken}, {Rob}, {Rolland}, {Rosier-Lees}, {Rowell}, {Rudak},
  {Ruppel}, {Sahakian}, {Santangelo}, {Saug{\'e}}, {Schlenker}, {Schlickeiser},
  {Schr{\"o}der}, {Schwanke}, {Schwarzburg}, {Schwemmer}, {Shalchi}, {Sol},
  {Spangler}, {Stawarz}, {Steenkamp}, {Stegmann}, {Superina}, {Tam},
  {Tavernet}, {Terrier}, {van Eldik}, {Vasileiadis}, {Venter}, {Vialle},
  {Vincent}, {Vivier}, {V{\"o}lk}, {Volpe}, {Wagner}, {Ward}, \&
  {Zdziarski}}]{2007ApJ...664L..71A}
{Aharonian}, F., {Akhperjanian}, A.~G., {Bazer-Bachi}, A.~R., {et~al.} 2007,
  \apjl, 664, L71

\bibitem[{{Aleksi{\'c}} {et~al.}(2011){Aleksi{\'c}}, {Antonelli}, {Antoranz},
  {Backes}, {Barrio}, {Bastieri}, {Becerra Gonz{\'a}lez}, {Bednarek},
  {Berdyugin}, {Berger}, {Bernardini}, {Biland}, {Blanch}, {Bock}, {Boller},
  {Bonnoli}, {Borla Tridon}, {Braun}, {Bretz}, {Ca{\~n}ellas}, {Carmona},
  {Carosi}, {Colin}, {Colombo}, {Contreras}, {Cortina}, {Cossio}, {Covino},
  {Dazzi}, {De Angelis}, {De Cea del Pozo}, {De Lotto}, {Delgado Mendez},
  {Diago Ortega}, {Doert}, {Dom{\'{\i}}nguez}, {Dominis Prester}, {Dorner},
  {Doro}, {Elsaesser}, {Ferenc}, {Fonseca}, {Font}, {Fruck}, {Garc{\'{\i}}a
  L{\'o}pez}, {Garczarczyk}, {Garrido}, {Giavitto}, {Godinovi{\'c}}, {Hadasch},
  {H{\"a}fner}, {Herrero}, {Hildebrand}, {H{\"o}hne-M{\"o}nch}, {Hose},
  {Hrupec}, {Huber}, {Jogler}, {Klepser}, {Kr{\"a}henb{\"u}hl}, {Krause}, {La
  Barbera}, {Lelas}, {Leonardo}, {Lindfors}, {Lombardi}, {L{\'o}pez}, {Lorenz},
  {Makariev}, {Maneva}, {Mankuzhiyil}, {Mannheim}, {Maraschi}, {Mariotti},
  {Mart{\'{\i}}nez}, {Mazin}, {Meucci}, {Miranda}, {Mirzoyan}, {Miyamoto},
  {Mold{\'o}n}, {Moralejo}, {Nieto}, {Nilsson}, {Orito}, {Oya}, {Paneque},
  {Paoletti}, {Pardo}, {Paredes}, {Partini}, {Pasanen}, {Pauss},
  {Perez-Torres}, {Persic}, {Peruzzo}, {Pilia}, {Pochon}, {Prada}, {Prada
  Moroni}, {Prandini}, {Puljak}, {Reichardt}, {Reinthal}, {Rhode}, {Rib{\'o}},
  {Rico}, {R{\"u}gamer}, {Saggion}, {Saito}, {Saito}, {Salvati}, {Satalecka},
  {Scalzotto}, {Scapin}, {Schultz}, {Schweizer}, {Shayduk}, {Shore},
  {Sillanp{\"a}{\"a}}, {Sitarek}, {Sobczynska}, {Spanier}, {Spiro}, {Stamerra},
  {Steinke}, {Storz}, {Strah}, {Suri{\'c}}, {Takalo}, {Tavecchio}, {Temnikov},
  {Terzi{\'c}}, {Tescaro}, {Teshima}, {Thom}, {Tibolla}, {Torres}, {Treves},
  {Vankov}, {Vogler}, {Wagner}, {Weitzel}, {Zabalza}, {Zandanel}, {Zanin},
  {MAGIC Collaboration}, {Tanaka}, {Wood}, \&
  {Buson}}]{2011ApJ...730L...8Aleksic}
{Aleksi{\'c}}, J., {Antonelli}, L.~A., {Antoranz}, P., {et~al.} 2011, \apjl,
  730, L8

\bibitem[{{Aller} {et~al.}(1985){Aller}, {Aller}, \&
  {Hughes}}]{1985ApJ...298..296A}
{Aller}, H.~D., {Aller}, M.~F., \& {Hughes}, P.~A. 1985, \apj, 298, 296

\bibitem[{{Atwood} {et~al.}(2009){Atwood}, {Abdo}, {Ackermann}, {Althouse},
  {Anderson}, {Axelsson}, {Baldini}, {Ballet}, {Band}, {Barbiellini}, \&
  et~al.}]{2009ApJ...697.1071A}
{Atwood}, W.~B., {Abdo}, A.~A., {Ackermann}, M., {et~al.} 2009, \apj, 697, 1071

\bibitem[{{Begelman} {et~al.}(1984){Begelman}, {Blandford}, \&
  {Rees}}]{Begelman84}
{Begelman}, M.~C., {Blandford}, R.~D., \& {Rees}, M.~J. 1984, Reviews of Modern
  Physics, 56, 255

\bibitem[{{Bevington} \& {Robinson}(2003)}]{2003drea.book.....Bevington}
{Bevington}, P.~R., \& {Robinson}, D.~K. 2003, {Data reduction and error
  analysis for the physical sciences} (3rd ed., by Philip R.~Bevington, and
  Keith D.~Robinson.~Boston, MA: McGraw-Hill, ISBN 0-07-247227-8, 2003)

\bibitem[{{Bhatta} {et~al.}(2016){Bhatta}, {Zola}, {Stawarz}, {Ostrowski},
  {Winiarski}, {Og{\l}oza}, {Dr{\'o}{\.z}d{\.z}}, {Siwak}, {Liakos},
  {Kozie{\l}-Wierzbowska}, {Gazeas}, {Debski}, {Kundera}, {Stachowski}, \&
  {Paliya}}]{2016ApJ...832...47B}
{Bhatta}, G., {Zola}, S., {Stawarz}, {\L}., {et~al.} 2016, \apj, 832, 47

\bibitem[{{B{\"o}ttcher} {et~al.}(2013){B{\"o}ttcher}, {Reimer}, {Sweeney}, \&
  {Prakash}}]{2013ApJ...768...54B}
{B{\"o}ttcher}, M., {Reimer}, A., {Sweeney}, K., \& {Prakash}, A. 2013, \apj,
  768, 54

\bibitem[{{Burnham} \& {Anderson}(2004)}]{Burnham04}
{Burnham}, K.~P., \& {Anderson}, D.~R. 2004, in New York : Springer, Vol.~54,
  Physics of Active Galaxies, ed. G.~V. {Bicknell}, M.~A. {Dopita}, \& P.~J.
  {Quinn}, 91

\bibitem[{{de Young}(2002)}]{DeYoung02}
{de Young}, D.~S. 2002, {The physics of extragalactic radio sources}
  (University of Chicago Press, 2002)

\bibitem[{{Deeming}(1975)}]{Deeming75}
{Deeming}, T.~J. 1975, \apss, 36, 137

\bibitem[{{Edelson} {et~al.}(2013){Edelson}, {Mushotzky}, {Vaughan}, {Scargle},
  {Gandhi}, {Malkan}, \& {Baumgartner}}]{Edelson13}
{Edelson}, R., {Mushotzky}, R., {Vaughan}, S., {et~al.} 2013, \apj, 766, 16

\bibitem[{{Edelson} \& {Krolik}(1988)}]{Edelson88}
{Edelson}, R.~A., \& {Krolik}, J.~H. 1988, \apj, 333, 646

\bibitem[{{Emmanoulopoulos} {et~al.}(2013){Emmanoulopoulos}, {McHardy}, \&
  {Papadakis}}]{Emmanoulopoulos13}
{Emmanoulopoulos}, D., {McHardy}, I.~M., \& {Papadakis}, I.~E. 2013, \mnras,
  433, 907

\bibitem[{{Falomo} {et~al.}(2014){Falomo}, {Pian}, \&
  {Treves}}]{2014A&ARv..22...73F}
{Falomo}, R., {Pian}, E., \& {Treves}, A. 2014, \aapr, 22, 73

\bibitem[{{Finke} \& {Becker}(2015)}]{2015ApJ...809...85F}
{Finke}, J.~D., \& {Becker}, P.~A. 2015, \apj, 809, 85

\bibitem[{{Fiorucci} \& {Tosti}(1996)}]{1996A&AS..116..403F}
{Fiorucci}, M., \& {Tosti}, G. 1996, \aaps, 116, 403

\bibitem[{{Foschini} {et~al.}(2011){Foschini}, {Ghisellini}, {Tavecchio},
  {Bonnoli}, \& {Stamerra}}]{Foschini11}
{Foschini}, L., {Ghisellini}, G., {Tavecchio}, F., {Bonnoli}, G., \&
  {Stamerra}, A. 2011, \aap, 530, A77

\bibitem[{{Gehrels} {et~al.}(2004){Gehrels}, {Chincarini}, {Giommi}, {Mason},
  {Nousek}, {Wells}, {White}, {Barthelmy}, {Burrows}, {Cominsky}, {Hurley},
  {Marshall}, {M{\'e}sz{\'a}ros}, {Roming}, {Angelini}, {Barbier}, {Belloni},
  {Campana}, {Caraveo}, {Chester}, {Citterio}, {Cline}, {Cropper}, {Cummings},
  {Dean}, {Feigelson}, {Fenimore}, {Frail}, {Fruchter}, {Garmire}, {Gendreau},
  {Ghisellini}, {Greiner}, {Hill}, {Hunsberger}, {Krimm}, {Kulkarni}, {Kumar},
  {Lebrun}, {Lloyd-Ronning}, {Markwardt}, {Mattson}, {Mushotzky}, {Norris},
  {Osborne}, {Paczynski}, {Palmer}, {Park}, {Parsons}, {Paul}, {Rees},
  {Reynolds}, {Rhoads}, {Sasseen}, {Schaefer}, {Short}, {Smale}, {Smith},
  {Stella}, {Tagliaferri}, {Takahashi}, {Tashiro}, {Townsley}, {Tueller},
  {Turner}, {Vietri}, {Voges}, {Ward}, {Willingale}, {Zerbi}, \&
  {Zhang}}]{2004ApJ...611.1005G}
{Gehrels}, N., {Chincarini}, G., {Giommi}, P., {et~al.} 2004, \apj, 611, 1005

\bibitem[{{Gelman} \& {Rubin}(1992)}]{Gelman92}
{Gelman}, A., \& {Rubin}, D.~B. 1992, Statistical Science, 7, 457

\bibitem[{{Ghisellini} {et~al.}(1998){Ghisellini}, {Celotti}, {Fossati},
  {Maraschi}, \& {Comastri}}]{1998MNRAS.301..451G}
{Ghisellini}, G., {Celotti}, A., {Fossati}, G., {Maraschi}, L., \& {Comastri},
  A. 1998, \mnras, 301, 451

\bibitem[{{Glass}(1999)}]{1999hia..book.....G}
{Glass}, I.~S. 1999, {Handbook of Infrared Astronomy}, ed. R.~{Ellis},
  J.~{Huchra}, S.~{Kahn}, G.~{Rieke}, \& P.~B. {Stetson} (Cambridge University
  Press, Aug 13, 1999, page 63)

\bibitem[{{Goyal} {et~al.}(2017){Goyal}, {Stawarz}, {Ostrowski}, {Larionov},
  {Gopal-Krishna}, {Wiita}, {Joshi}, {Soida}, \& {Agudo}}]{AG17}
{Goyal}, A., {Stawarz}, {\L}., {Ostrowski}, M., {et~al.} 2017, \apj, 837, 127

\bibitem[{{Howell} {et~al.}(2014){Howell}, {Sobeck}, {Haas}, {Still},
  {Barclay}, {Mullally}, {Troeltzsch}, {Aigrain}, {Bryson}, {Caldwell},
  {Chaplin}, {Cochran}, {Huber}, {Marcy}, {Miglio}, {Najita}, {Smith},
  {Twicken}, \& {Fortney}}]{Howell14}
{Howell}, S.~B., {Sobeck}, C., {Haas}, M., {et~al.} 2014, \pasp, 126, 398

\bibitem[{{Hudec} {et~al.}(2013){Hudec}, {Ba{\v s}ta}, {Pihajoki}, \&
  {Valtonen}}]{Hudec13}
{Hudec}, R., {Ba{\v s}ta}, M., {Pihajoki}, P., \& {Valtonen}, M. 2013, \aap,
  559, A20

\bibitem[{{Hurvich} \& {Tsai}(1989)}]{Hurvich89}
{Hurvich}, C.~M., \& {Tsai}, C.~L. 1989, Biometrika, 76, 297

\bibitem[{{Isobe} {et~al.}(2015){Isobe}, {Sato}, {Ueda}, {Hayashida},
  {Shidatsu}, {Kawamuro}, {Ueno}, {Sugizaki}, {Sugimoto}, {Mihara}, {Matsuoka},
  \& {Negoro}}]{2015ApJ...798...27I}
{Isobe}, N., {Sato}, R., {Ueda}, Y., {et~al.} 2015, \apj, 798, 27

\bibitem[{{Kastendieck} {et~al.}(2011){Kastendieck}, {Ashley}, \&
  {Horns}}]{2011A&A...531A.123K}
{Kastendieck}, M.~A., {Ashley}, M.~C.~B., \& {Horns}, D. 2011, \aap, 531, A123

\bibitem[{{Kataoka} {et~al.}(2001){Kataoka}, {Takahashi}, {Wagner}, {Iyomoto},
  {Edwards}, {Hayashida}, {Inoue}, {Madejski}, {Takahara}, {Tanihata}, \&
  {Kawai}}]{2001ApJ...560..659K}
{Kataoka}, J., {Takahashi}, T., {Wagner}, S.~J., {et~al.} 2001, \apj, 560, 659

\bibitem[{{Kelly} {et~al.}(2009){Kelly}, {Bechtold}, \&
  {Siemiginowska}}]{Kelly09}
{Kelly}, B.~C., {Bechtold}, J., \& {Siemiginowska}, A. 2009, \apj, 698, 895

\bibitem[{{Kelly} {et~al.}(2014){Kelly}, {Becker}, {Sobolewska},
  {Siemiginowska}, \& {Uttley}}]{Kelly14}
{Kelly}, B.~C., {Becker}, A.~C., {Sobolewska}, M., {Siemiginowska}, A., \&
  {Uttley}, P. 2014, \apj, 788, 33

\bibitem[{{Kelly} {et~al.}(2011){Kelly}, {Sobolewska}, \&
  {Siemiginowska}}]{Kelly11}
{Kelly}, B.~C., {Sobolewska}, M., \& {Siemiginowska}, A. 2011, \apj, 730, 52

\bibitem[{{Kushwaha} {et~al.}(2016){Kushwaha}, {Chandra}, {Misra},
  {Sahayanathan}, {Singh}, \& {Baliyan}}]{Kushwaha16}
{Kushwaha}, P., {Chandra}, S., {Misra}, R., {et~al.} 2016, \apjl, 822, L13

\bibitem[{{Liodakis} {et~al.}(2017){Liodakis}, {Pavlidou}, {Hovatta},
  {Max-Moerbeck}, {Pearson}, {Richards}, \& {Readhead}}]{Liodakis17}
{Liodakis}, I., {Pavlidou}, V., {Hovatta}, T., {et~al.} 2017, \mnras, 467, 4565

\bibitem[{{Lister} {et~al.}(2016){Lister}, {Aller}, {Aller}, {Homan},
  {Kellermann}, {Kovalev}, {Pushkarev}, {Richards}, {Ros}, \&
  {Savolainen}}]{2016AJ....152...12L}
{Lister}, M.~L., {Aller}, M.~F., {Aller}, H.~D., {et~al.} 2016, \aj, 152, 12

\bibitem[{{Lomb}(1976)}]{Lomb76}
{Lomb}, N.~R. 1976, \apss, 39, 447

\bibitem[{{Massaro} {et~al.}(2003){Massaro}, {Giommi}, {Perri}, {Tagliaferri},
  {Nesci}, {Tosti}, {Ciprini}, {Montagni}, {Ravasio}, {Ghisellini}, {Frasca},
  {Marilli}, {Valentini}, {Kurtanidze}, \& {Nikolashvili}}]{Massaro03}
{Massaro}, E., {Giommi}, P., {Perri}, M., {et~al.} 2003, \aap, 399, 33

\bibitem[{{McHardy} {et~al.}(2006){McHardy}, {Koerding}, {Knigge}, {Uttley}, \&
  {Fender}}]{Mchardy06}
{McHardy}, I.~M., {Koerding}, E., {Knigge}, C., {Uttley}, P., \& {Fender},
  R.~P. 2006, \nat, 444, 730

\bibitem[{{Meier}(2012)}]{Meier12}
{Meier}, D.~L. 2012, {Black Hole Astrophysics: The Engine Paradigm} (Springer,
  Verlag Berlin Heidelberg, 2012)

\bibitem[{{Nilsson} {et~al.}(2010){Nilsson}, {Takalo}, {Lehto}, \&
  {Sillanp{\"a}{\"a}}}]{Nilsson2010}
{Nilsson}, K., {Takalo}, L.~O., {Lehto}, H.~J., \& {Sillanp{\"a}{\"a}}, A.
  2010, \aap, 516, A60

\bibitem[{{O'Brien}(2017)}]{OBrien17}
{O'Brien}, S. 2017, ArXiv e-prints, arXiv:1708.02160

\bibitem[{{O'Riordan} {et~al.}(2017){O'Riordan}, {Pe'er}, \&
  {McKinney}}]{2017ApJ...843...81O}
{O'Riordan}, M., {Pe'er}, A., \& {McKinney}, J.~C. 2017, \apj, 843, 81

\bibitem[{{Qian} \& {Tao}(2003)}]{2003PASP..115..490Q}
{Qian}, B., \& {Tao}, J. 2003, \pasp, 115, 490

\bibitem[{{Rani} {et~al.}(2013){Rani}, {Lott}, {Krichbaum}, {Fuhrmann}, \&
  {Zensus}}]{Rani13}
{Rani}, B., {Lott}, B., {Krichbaum}, T.~P., {Fuhrmann}, L., \& {Zensus}, J.~A.
  2013, \aap, 557, A71

\bibitem[{{Revalski} {et~al.}(2014){Revalski}, {Nowak}, {Wiita}, {Wehrle}, \&
  {Unwin}}]{2014ApJ...785...60R}
{Revalski}, M., {Nowak}, D., {Wiita}, P.~J., {Wehrle}, A.~E., \& {Unwin}, S.~C.
  2014, \apj, 785, 60

\bibitem[{{Richards} {et~al.}(2011){Richards}, {Max-Moerbeck}, {Pavlidou},
  {King}, {Pearson}, {Readhead}, {Reeves}, {Shepherd}, {Stevenson},
  {Weintraub}, {Fuhrmann}, {Angelakis}, {Zensus}, {Healey}, {Romani}, {Shaw},
  {Grainge}, {Birkinshaw}, {Lancaster}, {Worrall}, {Taylor}, {Cotter}, \&
  {Bustos}}]{2011ApJS..194...29R}
{Richards}, J.~L., {Max-Moerbeck}, W., {Pavlidou}, V., {et~al.} 2011, \apjs,
  194, 29

\bibitem[{{Saito} {et~al.}(2013){Saito}, {Stawarz}, {Tanaka}, {Takahashi},
  {Madejski}, \& {D Ammando}}]{Saito13}
{Saito}, S., {Stawarz}, {\L}., {Tanaka}, Y.~T., {et~al.} 2013, \apjl, 766, L11

\bibitem[{{Sandrinelli} {et~al.}(2016){Sandrinelli}, {Covino}, {Dotti}, \&
  {Treves}}]{2016AJ....151...54S}
{Sandrinelli}, A., {Covino}, S., {Dotti}, M., \& {Treves}, A. 2016, \aj, 151,
  54

\bibitem[{{Scargle}(1982)}]{Scargle82}
{Scargle}, J.~D. 1982, \apj, 263, 835

\bibitem[{{Seta} {et~al.}(2009){Seta}, {Isobe}, {Tashiro}, {Yaji}, {Arai},
  {Fukuhara}, {Kohno}, {Nakanishi}, {Sasada}, {Shimajiri}, {Tosaki}, {Uemura},
  {Anderhub}, {Antonelli}, {Antoranz}, {Backes}, {Baixeras}, {Balestra},
  {Barrio}, {Bastieri}, {Becerra Gonz{\'a}lez}, {Becker}, {Bednarek}, {Berger},
  {Bernardini}, {Biland}, {Bock}, {Bonnoli}, {Bordas}, {Borla Tridon},
  {Bosch-Ramon}, {Bose}, {Braun}, {Bretz}, {Britvitch}, {Camara}, {Carmona},
  {Commichau}, {Contreras}, {Cortina}, {Costado Dios}, {Covino}, {Curtef},
  {Dazzi}, {de Angelis}, {de Cea Del Pozo}, {de Los Reyes}, {de Lotto}, {de
  Maria}, {de Sabata}, {Delgado M{\'e}ndez}, {Dom{\'{\i}}nguez}, {Dorner},
  {Doro}, {Elsaesser}, {Errando}, {Ferenc}, {Fern{\'a}ndez}, {Firpo},
  {Fonseca}, {Font}, {Galante}, {Garc{\'{\i}}a L{\'o}pez}, {Garczarczyk},
  {Gaug}, {Goebel}, {Hadasch}, {Hayashida}, {Herrero}, {Hildebrand},
  {H{\"o}hne-M{\"o}nch}, {Hose}, {Hsu}, {Jogler}, {Kranich}, {La Barbera},
  {Laille}, {Leonardo}, {Lindfors}, {Lombardi}, {Longo}, {L{\'o}pez}, {Lorenz},
  {Majumdar}, {Maneva}, {Mankuzhiyil}, {Mannheim}, {Maraschi}, {Mariotti},
  {Mart{\'{\i}}nez}, {Mazin}, {Meucci}, {Meyer}, {Miguel Miranda}, {Mirzoyan},
  {Miyamoto}, {Mold{\'o}n}, {Moles}, {Moralejo}, {Nieto}, {Nilsson},
  {Ninkovic}, {Otte}, {Oya}, {Paoletti}, {Paredes}, {Pasanen}, {Pascoli},
  {Pauss}, {Pegna}, {Perez-Torres}, {Persic}, {Peruzzo}, {Prada}, {Prandini},
  {Puchades}, {Reichardt}, {Rhode}, {Rib{\'o}}, {Rico}, {Rissi}, {Robert},
  {R{\"u}gamer}, {Saggion}, {Saito}, {Salvati}, {S{\'a}nchez-Conde},
  {Satalecka}, {Scalzotto}, {Scapin}, {Schweizer}, {Shayduk}, {Shore}, {Sidro},
  {Sierpowska-Bartosik}, {Sillanp{\"a}{\"a}}, {Sitarek}, {Sobczynska},
  {Spanier}, {Stamerra}, {Stark Schneebeli}, {Takalo}, {Tavecchio}, {Temnikov},
  {Tescaro}, {Teshima}, {Tluczykont}, {Torres}, {Turini}, {Vankov}, {Wagner},
  {Wittek}, {Zabalza}, {Zandanel}, {Zanin}, \&
  {Zapatero}}]{2009PASJ...61.1011S}
{Seta}, H., {Isobe}, N., {Tashiro}, M.~S., {et~al.} 2009, \pasj, 61, 1011

\bibitem[{{Sillanpaa} {et~al.}(1996){Sillanpaa}, {Takalo}, {Pursimo}, {Lehto},
  {Nilsson}, {Teerikorpi}, {Heinaemaeki}, {Kidger}, {de Diego},
  {Gonzalez-Perez}, {Rodriguez-Espinosa}, {Mahoney}, {Boltwood},
  {Dultzin-Hacyan}, {Benitez}, {Turner}, {Robertson}, {Honeycut}, {Efimov},
  {Shakhovskoy}, {Charles}, {Schramm}, {Borgeest}, {Linde}, {Weneit}, {Kuehl},
  {Schramm}, {Sadun}, {Grashuis}, {Heidt}, {Wagner}, {Bock}, {Kuemmel},
  {Heines}, {Fiorucci}, {Tosti}, {Ghisellini}, {Raiteri}, {Villata}, {de
  Francesco}, {Bosio}, \& {Latini}}]{1996A&A...305L..17S}
{Sillanpaa}, A., {Takalo}, L.~O., {Pursimo}, T., {et~al.} 1996, \aap, 305, L17

\bibitem[{{Sobolewska} {et~al.}(2014){Sobolewska}, {Siemiginowska}, {Kelly}, \&
  {Nalewajko}}]{2014ApJ...786..143S}
{Sobolewska}, M.~A., {Siemiginowska}, A., {Kelly}, B.~C., \& {Nalewajko}, K.
  2014, \apj, 786, 143

\bibitem[{{Takalo}(1994)}]{Takalo94}
{Takalo}, L.~O. 1994, Vistas in Astronomy, 38, 77

\bibitem[{{Takalo} {et~al.}(1994){Takalo}, {Sillanpaeae}, \&
  {Nilsson}}]{1994A&AS..107..497T}
{Takalo}, L.~O., {Sillanpaeae}, A., \& {Nilsson}, K. 1994, \aaps, 107

\bibitem[{{Timmer} \& {Koenig}(1995)}]{TK95}
{Timmer}, J., \& {Koenig}, M. 1995, \aap, 300, 707

\bibitem[{{Ulrich} {et~al.}(1997){Ulrich}, {Maraschi}, \& {Urry}}]{Ulrich97}
{Ulrich}, M.-H., {Maraschi}, L., \& {Urry}, C.~M. 1997, \araa, 35, 445

\bibitem[{{Urry} \& {Padovani}(1995)}]{1995PASP..107..803Urry}
{Urry}, C.~M., \& {Padovani}, P. 1995, \pasp, 107, 803

\bibitem[{{Uttley} {et~al.}(2002){Uttley}, {McHardy}, \&
  {Papadakis}}]{Uttley02}
{Uttley}, P., {McHardy}, I.~M., \& {Papadakis}, I.~E. 2002, \mnras, 332, 231

\bibitem[{{Valtaoja} {et~al.}(2000){Valtaoja}, {Ter{\"a}sranta}, {Tornikoski},
  {Sillanp{\"a}{\"a}}, {Aller}, {Aller}, \& {Hughes}}]{Valtaoja00}
{Valtaoja}, E., {Ter{\"a}sranta}, H., {Tornikoski}, M., {et~al.} 2000, \apj,
  531, 744

\bibitem[{{Valtonen} \& {Sillanp{\"a}{\"a}}(2011)}]{Valtonen11}
{Valtonen}, M., \& {Sillanp{\"a}{\"a}}, A. 2011, Acta Polytechnica, 51, 76

\bibitem[{{Valtonen} {et~al.}(2016){Valtonen}, {Zola}, {Ciprini}, {Gopakumar},
  {Matsumoto}, {Sadakane}, {Kidger}, {Gazeas}, {Nilsson}, {Berdyugin},
  {Piirola}, {Jermak}, {Baliyan}, {Alicavus}, {Boyd}, {Campas Torrent},
  {Campos}, {Carrillo G{\'o}mez}, {Caton}, {Chavushyan}, {Dalessio}, {Debski},
  {Dimitrov}, {Drozdz}, {Er}, {Erdem}, {Escartin P{\'e}rez}, {Fallah Ramazani},
  {Filippenko}, {Ganesh}, {Garcia}, {G{\'o}mez Pinilla}, {Gopinathan},
  {Haislip}, {Hudec}, {Hurst}, {Ivarsen}, {Jelinek}, {Joshi}, {Kagitani},
  {Kaur}, {Keel}, {LaCluyze}, {Lee}, {Lindfors}, {Lozano de Haro}, {Moore},
  {Mugrauer}, {Naves Nogues}, {Neely}, {Nelson}, {Ogloza}, {Okano}, {Pandey},
  {Perri}, {Pihajoki}, {Poyner}, {Provencal}, {Pursimo}, {Raj}, {Reichart},
  {Reinthal}, {Sadegi}, {Sakanoi}, {Salto Gonz{\'a}lez}, {Sameer}, {Schweyer},
  {Siwak}, {Sold{\'a}n Alfaro}, {Sonbas}, {Steele}, {Stocke}, {Strobl},
  {Takalo}, {Tomov}, {Tremosa Espasa}, {Valdes}, {Valero P{\'e}rez},
  {Verrecchia}, {Webb}, {Yoneda}, {Zejmo}, {Zheng}, {Telting}, {Saario},
  {Reynolds}, {Kvammen}, {Gafton}, {Karjalainen}, {Harmanen}, \&
  {Blay}}]{Valtonen16}
{Valtonen}, M.~J., {Zola}, S., {Ciprini}, S., {et~al.} 2016, \apjl, 819, L37

\bibitem[{{VanderPlas}(2018)}]{VanderPlas18} {VanderPlas}, J.~T.\ 2018, \apjs, 236, 16 

\bibitem[{{Vaughan} {et~al.}(2016){Vaughan}, {Uttley}, {Markowitz},
  {Huppenkothen}, {Middleton}, {Alston}, {Scargle}, \& {Farr}}]{Vaughan16}
{Vaughan}, S., {Uttley}, P., {Markowitz}, A.~G., {et~al.} 2016, \mnras, 461,
  3145

\bibitem[{{Villforth} {et~al.}(2010){Villforth}, {Nilsson}, {Heidt}, {Takalo},
  {Pursimo}, {Berdyugin}, {Lindfors}, {Pasanen}, {Winiarski}, {Drozdz},
  {Ogloza}, {Kurpinska-Winiarska}, {Siwak}, {Koziel-Wierzbowska}, {Porowski},
  {Kuzmicz}, {Krzesinski}, {Kundera}, {Wu}, {Zhou}, {Efimov}, {Sadakane},
  {Kamada}, {Ohlert}, {Hentunen}, {Nissinen}, {Dietrich}, {Assef}, {Atlee},
  {Bird}, {Depoy}, {Eastman}, {Peeples}, {Prieto}, {Watson}, {Yee}, {Liakos},
  {Niarchos}, {Gazeas}, {Dogru}, {Donmez}, {Marchev}, {Coggins-Hill},
  {Mattingly}, {Keel}, {Haque}, {Aungwerojwit}, \&
  {Bergvall}}]{2010MNRAS.402.2087V}
{Villforth}, C., {Nilsson}, K., {Heidt}, J., {et~al.} 2010, \mnras, 402, 2087

\bibitem[{{Wagner} \& {Witzel}(1995)}]{1995ARA&A..33..163W}
{Wagner}, S.~J., \& {Witzel}, A. 1995, \araa, 33, 163

\bibitem[{{Wills} {et~al.}(2011){Wills}, {Wills}, \&
  {Breger}}]{2011ApJS..194...19W}
{Wills}, B.~J., {Wills}, D., \& {Breger}, M. 2011, \apjs, 194, 19

\bibitem[{{Wills} {et~al.}(1992){Wills}, {Wills}, {Breger}, {Antonucci}, \&
  {Barvainis}}]{1992ApJ...398..454Wills}
{Wills}, B.~J., {Wills}, D., {Breger}, M., {Antonucci}, R.~R.~J., \&
  {Barvainis}, R. 1992, \apj, 398, 454

\bibitem[{{Zola} {et~al.}(2016){Zola}, {Valtonen}, {Bhatta}, {Goyal}, {Debski},
  {Baran}, {Krzesinski}, {Siwak}, {Ciprini}, {Gopakumar}, {Jermak}, {Nilsson},
  {Reichart}, {Matsumoto}, {Sadakane}, {Gazeas}, {Kidger}, {Piirola},
  {Alicavus}, {Baliyan}, {Berdyugin}, {Boyd}, {Campas Torrent}, {Campos},
  {Carrillo G{\'o}mez}, {Caton}, {Chavushyan}, {Dalessio}, {Dimitrov},
  {Drozdz}, {Er}, {Erdem}, {Escartin P{\'e}rez}, {Fallah Ramazani},
  {Filippenko}, {Garcia}, {G{\'o}mez Pinilla}, {Gopinathan}, {Haislip},
  {Harmanen}, {Hudec}, {Hurst}, {Ivarsen}, {Jelinek}, {Joshi}, {Kagitani},
  {Kaur}, {Keel}, {LaCluyze}, {Lee}, {Lindfors}, {Lozano de Haro}, {Moore},
  {Mugrauer}, {Naves Nogues}, {Neely}, {Nelson}, {Ogloza}, {Okano}, {Pandey},
  {Perri}, {Pihajoki}, {Poyner}, {Provencal}, {Pursimo}, {Raj}, {Reinthal},
  {Sadegi}, {Sakanoi}, {Sameer}, {Salto Gonz{\'a}lez}, {Schweyer}, {Sold{\'a}n
  Alfaro}, {Karaman}, {Sonbas}, {Steele}, {Stocke}, {Strobl}, {Takalo},
  {Tomov}, {Tremosa Espasa}, {Valdes}, {Valero P{\'e}rez}, {Verrecchia},
  {Webb}, {Yoneda}, {Zejmo}, {Zheng}, {Telting}, {Saario}, {Reynolds},
  {Kvammen}, {Gafton}, {Karjalainen}, \& {Blay}}]{Zola16}
{Zola}, S., {Valtonen}, M., {Bhatta}, G., {et~al.} 2016, Galaxies, 4, 41

\end{thebibliography}
\end{document}